\documentclass[fleqn,usenatbib]{mnras}

\usepackage[T1]{fontenc}
\usepackage{ae,aecompl}

\usepackage[utf8]{inputenc}

\usepackage{float}
\usepackage{amsmath,amssymb}
\usepackage{mathrsfs}
\usepackage{theorem}
\usepackage{graphicx}
\usepackage{caption}
\usepackage{subcaption}
\usepackage{color}
\usepackage{amsfonts}
\usepackage{wasysym}
\usepackage{hyperref} 
\usepackage{mathtools}
\usepackage{dcolumn}

\newtheorem{teo}{Theorem}[section]
\newtheorem{defi}{Definition}[section]

\definecolor{blue}{rgb}{0,0,1}
\definecolor{green}{rgb}{0,0.65,0.5}
\definecolor{verde}{rgb}{0.,.5,0.4}
\definecolor{marron}{rgb}{0.7,0.2,0.1}
\definecolor{red}{rgb}{1,0,0}
\definecolor{vio}{rgb}{0.66,0,1}
\definecolor{ama}{rgb}{1,1,0}
\definecolor{veroscuro}{rgb}{0.3,0.36,0.33}

\title[Gravitational lens optical scalars...]{Gravitational lens optical scalars in terms of energy-momentum distributions 
	in the cosmological framework.}

\author[E.F.Boero and O.M.Moreschi]{
Ezequiel F. Boero,$^{1}$\thanks{E-mail:boero@famaf.unc.edu.ar}
Osvaldo M. Moreschi,$^{2,3}$\thanks{E-mail:o.moreschi@unc.edu.ar}
\\
$^{1}$ Instituto de Astronom\'\i{}a Te\'{o}rica y Experimental (IATE), CONICET,
Observatorio Astron\'{o}mico de Córdoba,\\
Laprida 854, (X5000BGR) C\'{o}rdoba, Argentina.\\
$^{2}$Facultad de Matem\'{a}tica, Astronom\'\i{}a, F\'\i{}sica y Computaci\'{o}n (FaMAF),
Universidad Nacional de C\'{o}rdoba, \\ 
Medina Allende S/N, X5000HUA, Córdoba, Argentina.  \\
$^{3}$Instituto de F\'\i{}sica Enrique Gaviola, IFEG, CONICET,Medina Allende S/N, X5000HUA, Córdoba, Argentina. 
}

\date{Accepted XXX. Received YYY; in original form ZZZ}

\pubyear{2017}

\begin{document}
\label{firstpage}
\pagerange{\pageref{firstpage}--\pageref{lastpage}}
\maketitle

\begin{abstract}
	
We present new results on gravitational lensing over cosmological Robertson-Walker 
backgrounds which extend and generalize previous works. 
Our expressions show the presence of new terms and factors which have been 
neglected in the literature on the subject. 
The new equations derived here for the optical scalars allow to deal with more 
general matter content including sources with non Newtonian components of the
energy-momentum tensor and arbitrary motion. 
Our treatment is within the framework of weak gravitational
lenses in which first order effects of the curvature are considered.	
We have been able to make all calculations without
referring to the concept of deviation angle. This in turn,
makes the presentation shorter but also allows for the
consideration of global effects on the Robertson-Walker
background that have been neglected in the literature.

We also discuss two intensity magnifications, that we define in this article;
one coming from a natural geometrical construction in terms of the affine distance,
that we here call $\tilde{\mu}$, 
and the other adapted to cosmological discussions in terms of 
the redshift, that we call $\mu'$. 
We show that the natural intensity magnification $\tilde{\mu}$ 
coincides with the standard angular magnification 
($\mu$).
\end{abstract}

\begin{keywords}
gravitational lensing: weak -- gravitation -- cosmology: observations
\end{keywords}

%%%%%%%%%%%%%%%%%%%%%%%%%%%%%%%%%%%%%%%%%%%%%%%%%%

%%%%%%%%%%%%%%%%% BODY OF PAPER %%%%%%%%%%%%%%%%%%

\section{Introduction}

This work generalizes the results of reference \cite{Gallo11} to the cosmological framework
in which a Robertson-Walker (R-W) geometry dominates the large scale structure of the spacetime.
Therefore, we present new expressions for the lens optical scalars explicitly
in terms of the local curvature associated to compact lenses
in a cosmological scenario;
and for the case of spherically symmetric lenses, 
we express the scalars directly in terms of the energy-momentum tensor distribution.
We make no a priori assumptions on the nature of the lens, 
so that the new equations we present here admits a very general 
energy-matter content. 
In particular, they
show the contribution from components of the matter distribution that have been neglected 
previously.

The Universe is filled with objects and systems in motion, like
black holes that have received a kick after a coalescence of a binary 
system\cite{Tichy07,Baker08,Civano:2012bu,Gerosa:2016vip},
galaxies in a cluster\cite{Lokas:2003ks},
moving voids\cite{Lambas:2015afa}, etc.
It is therefore important to have at hand a formalism that allows to study
systems of gravitational lenses with motion.
We therefore have made no assumption on the nature of the motion of the lenses
and present new general expressions for the observable optical gravitational
lens scalars, that improve on previous works on the subject of moving gravitational
lens\cite{Kopeikin:1999ev,Frittelli:2002yx,Wucknitz:2004tw}.

We also present other new results including 
the relation between the intensity magnifications, that we define here,
with the standard angular magnification.

In our previous work\cite{Gallo11} on gravitational lenses, we have
presented new general expressions for the bending angle and the optical scalars 
for a congruence of null geodesics in the regime of weak gravitational lensing
over a flat background.
These formulas have the advantage of being explicitly gauge invariant and 
allow us to include more general forms of matter distributions than those 
discussed in most standard references on the subject\cite{Schneider92, Seitz94, 
	Wambsganss98, Bartelmann10}.

Systems involving gravitational lens effects appear naturally in
the cosmological context. And although the standard works on gravitational
lens, based on a flat background, are useful; new subtle issues arise
when the gravitational lens is studied in the cosmological framework.
We improve on the tools
shown in \cite{Gallo11}, and so provide with new expressions,
for weak gravitational lenses, that are
useful for the study of the matter content in cosmological astrophysical
systems.
Indeed, the expressions presented below are a tool to study in more detail
the missing mass problem; since we present equations describing 
gravitational lensing that describe more general situations;
enabling the characterization of more general energy momentum tensors,
in contrast to the usual Newtonian description of the dark
matter phenomena. 
Our expressions do not neglect spacelike 
components of the energy momentum tensor.
In fact, we have seen in the past that there is some suggestive evidence that the inclusion 
of the up to now neglected spacelike components of the energy momentum tensor turns 
out to be important for the study of dark matter phenomena as it was shown in 
\cite{Gallo:2011hi} where a simplified model with this features allows for excellent 
description of the matter content in clusters of galaxies, and for the velocity
rotation curves in galaxies.

It is probably worthwhile to mention that all the astrophysical observations
of the gravitational lens optical scalars, namely 
of the expansion and the shear, give results whose values are much less
than the unit value. For this reason one is convinced that 
pursuing calculations of weak gravitational lenses taking
into account the linear effects of the curvature is enough
to explain the observational data.	
In any case our starting point will be the exact geodesic
deviation equation; which happened to be linear in the curvature.
%\verde{
To be more precise, below we will present a decomposition of the curvature
in terms of a background term, that we will call $Q_B$, and an extra
term, that will call $Q_L$. By weak gravitational lens effects
we mean those that are deduced from the geodesic deviation equation at first order in the 
extra terms of the curvature fields: namely, $Q_L$.
In this decomposition of the total curvature
in terms of the background plus a lens term, it is not assumed
that the additional term must be small in any sense.
For example, $Q_L$ could be the fields that represent locally a
Schwarzschild black hole, or a 90\% deficit in the cosmological
density in the case of a void. That is we are not assuming at this
stage any kind of perturbation; it is just the representation
of an exact geometry, without perturbation, 
in terms of a decomposition with respect to a chosen background.
The details and implications of this statement will become
clear along the article.
However we advance that the effects of the background curvature
will be calculated exactly; so that our final equations will have 
quadratic terms of order $\mathscr{O}(Q_B \, Q_L)$.
%}

In this work, our main focus is to
broaden the range of validity of previous results 
and discussions to the case in which the background is within the family of 
Robertson-Walker geometries,
proceeding in a systematic way without introducing early assumptions of Newtonian 
character and further preconditions related to the observational configuration such as 
the well known approximation of thin lenses.
Regarding this last point,
here, instead we have chosen to consider such approximations after a general treatment of 
the exact equations governing the distortion of the images in the cosmological background.
The new improvements that we introduce require subtle but significant modifications in the 
treatment of the subject, that will be described along the article.
In particular, whenever appropriate we will mark the difference with the
simplistic approach to the subject in which all the physics of gravitational
lenses is encoded in the quotient of the projected surface mass density by
the so called critical mass density, normally denoted by 
$\Sigma/\Sigma_{cr}$\cite{Schneider92,Schneider06}. 
An interesting new result is for example the presence of a redshift factor correcting 
the widely used expressions of thin lenses as appears in eq. (51), part 1 of 
reference \cite{Schneider06} or eq. (16) of reference \cite{Wambsganss98}
(See section \ref{sec:thin-lenses}).
Our findings are relevant to works concerning tests of fundamental geometrical 
relations in observational cosmology which are based on the use of gravitational 
lensing; see for example \cite{Holanda:2015zpz, Liao:2015uzb}.

Let us mention here briefly other differences of our
approach to the subject of cosmic lenses with the standard references.
It has been suggested and is also widely believed that when describing
the gravitational lens optical scalars in the cosmological framework
one only needs to replace, in the flat background discussions, any
appearance of distance with the area distance, or angular diameter distance\cite{Schneider06};
we point out below that this imposes severe limitation on the class of
systems one can treat with this technique, and provide with the general
equations. 
In other occasions the presentation of gravitational lens physics relays
on the concept of effective local index of refraction\cite{Bartelmann:1999yn}
which can only be defined in very especial cases\cite{Schneider92}.
In contrast, our approach is to deduced the most general gravitational
lens equations that can be used in the cosmological framework.
In the process of obtaining these equations it is essential
to review the main concepts involved in the problem of gravitational lens
so that gauge invariant quantities can be derived without the use of
initial simplifying assumptions. For this reason we spend some time
below to make this review; which allows us for example to claim
that the only sensible notion of distance that one should use
when referring to observables in problems involving gravitational lenses
is the affine distance as defined by the observer. 
This does not contradict the use of area distance in calculations,
rather it is in contrast
to the claim that one should only use the angular diameter distance in
gravitational lens discussions in the cosmic framework.
We therefore emphasize along the article that the derived concepts
coming from gravitational lens effects concern always to comparisons
referring to same affine distances.
This approach lead us to define the concept of intensity magnification
that to our knowledge has not been used in the past.

In section \ref{sec:Grav-lens} we spent some time reviewing the basic concepts,
also with the intention to set the notation and language used in the rest of
this paper.
In section \ref{sec:FRW} we discuss null 
geodesics in R-W geometry and coordinates adapted to the observations.
Section \ref{sec:distances} contains a discussion of the different notions 
of distance appearing in cosmological studies together with the relation among them. 
We discuss different notions of magnifications in section \ref{sec:magnifications}.
Next, in section \ref{sec:RW-Lens} we present the expression for
the cosmological convergence, valid for arbitrary angular aperture and 
without approximations,
when the R-W geometry is considered as a gravitational lens.
The appearance of an additional lens over the R-W background is discussed in section 
\ref{sec:Addit-lens}; where our generalized expression are presented.
In this section the presentation deals with compact lenses of small 
extension in comparison with the cosmological dimension traversed by the photons; 
and the calculations of the
optical scalars are carried out within the framework of weak gravitational
lenses; in which first order effects of the curvature are used.
Section \ref{sec:Axial-symmetry} deals with the axially symmetric case.
In section \ref{sec:thin-lenses} we present the usual approximation of thin lenses;
while in section \ref{sec:esferica-lenses} the equations for a static spherically 
symmetric lens are written in its most general form.
Final comments are included  in section \ref{sec:final-comments}.
%\verde{
Lastly, in the appendix we include the geometrical structure of the 
cosmological background in terms of the less common presentation associated to
a null tetrad which is widely used in our approach.%}

\section{Gravitational lenses}\label{sec:Grav-lens}

\subsection{Preliminaries}

\begin{figure}%[H]
	\centering
	\includegraphics[clip,width=0.4\textwidth]{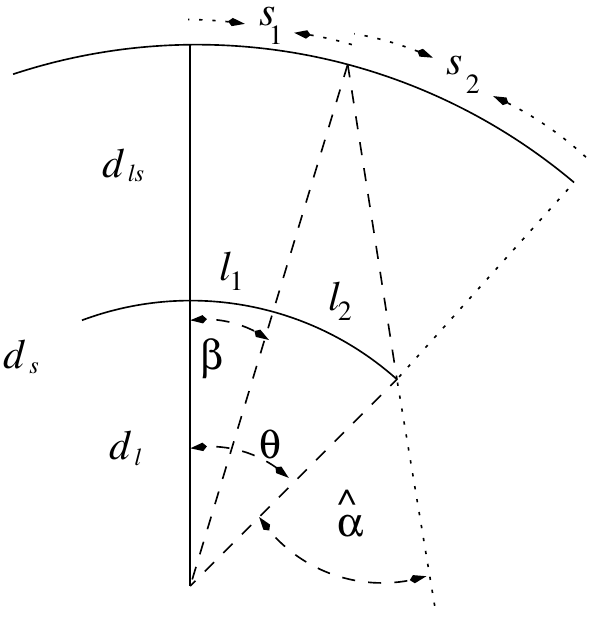}
	\caption{This graph shows the basic and familiar angular variables in terms of a simple
		flat background geometry.
		The letter $s$ denote sources, the letter $l$ denotes lens and the observer is
		assumed to be situated at the apex of the rays.}\label{f:fig0}
\end{figure}

The first thing to do is to agree in the language one would use to describe the 
phenomenology of gravitational lens effects.
The basic idea one has in mind is depicted in figure \ref{f:fig0};
where it is sketched the situation of an observer, at the bottom of the figure,
receiving light from a source at the furthest surface, which is affected
by a gravitational lens, along the path of the rays. 

In the discussion of the gravitational lens effects coming from an homogeneous
cosmological spacetime, one would not have a lens at a particular distance,
since the whole spacetime acts as a lens; however we will retain the
basic notions for the observed angle $\theta$ and \emph{would observe}
angle $\beta$; where the word `would' refers to what the astrophysicist 
would expect, if there where no gravitational lens effect, and therefore no curvature.

In any case, one can use the standard definitions of convergence and shear,
that comes from the relation of these angles.

Given a related pair of directions, represented by angles $(\beta,\theta)$, one can consider a small variation
from them, whose difference will be related by a linear relation. In the sphere of 
directions, they can be expressed as:
\begin{equation}\label{eq:standarlensequation}
\delta \beta^a = \mathcal{A}^a_{\; b}\,  \delta\theta^b,
\end{equation}
where the matrix $\mathcal{A}^a_{\; b}$ is in turn expressed by
\begin{equation}\label{eq:optical-scalars0}
\mathcal{A}^a_{\; b} =
\left( {\begin{array}{cc}
	1-\kappa-\gamma_1 & -\gamma_2  \\
	-\gamma_2& 1-\kappa+\gamma_1  \\
	\end{array} } \right);
\end{equation}
where the optical scalars $\kappa$, $\gamma_1$ and $\gamma_2$, are known as 
convergence $\kappa$ and shear components $\{\gamma_1,\gamma_2\}$, and have the information 
of distortion of the image of the source due to the lens effects.
In fact the shear is normally represented by the complex quantity
$\gamma_c = \gamma_1 + i \gamma_2$.

It should be emphasized that since the definition of $\delta \beta^a$ and 
$\delta\theta^b$ refer to two different geometries, one must say at what
distance is the source in both cases.
The calculations of the gravitational lens effects are carried out using
as fundamental variable the affine distance $\lambda$ from the point of 
observation\cite{Gallo11}, and with respect to the four velocity of
the observer. 
This point is also related to the notion of angular diameter distance, that we will
discuss again below for the case of homogeneous spacetimes; but here we will
only use the notion that a rod of length $\delta l$ which is observed to
subtend an angle $\delta \phi$ %;from which one 
defines the angular
diameter distance $D_A$ by:
\begin{equation}\label{eq:angular-diameter0}
\delta l = D_A \delta \phi .
\end{equation}
For each spacetime one has one notion of observed angular diameter distance.
In the above equation, one assumes the observation of some object of size $\delta l$,
so that one would have
\begin{equation}\label{eq:angular-diameter-beta}
\delta l = D_A(\eta) \delta \beta ;
\end{equation}
and
\begin{equation}\label{eq:angular-diameter-tita}
\delta l = D_A(g) \delta \theta ;
\end{equation}
where $\eta$ is the flat metric and $g$ the real physical metric.
In Minkowski spacetime the angular
diameter distance coincides with the affine distance, namely: $D_A(\eta)=\lambda$.
In a general spacetime, with metric $g$, one has the source at the
affine distance $\lambda_s$; therefore, in order to make the comparison in
equation (\ref{eq:standarlensequation}) one takes $\lambda_s = \lambda$.

\subsection{The null geodesic congruence}
Here we provide with a brief summary of the essentials equations needed to the 
treatment of gravitational lensing.

Let $u$ be an scalar function such that the level sets $u=$constant denote a 
family of null hypersurfaces; then we define the vector field $\ell^{a}$ from 
the one-form 
\begin{equation}\label{eq:ell_du}
\ell_{a} = \nabla_{a}u = (du)_a ;
\end{equation}
which satisfies
\begin{equation}
\ell^{a}\ell_{a} = 0, 
\end{equation}
and
\begin{equation}
\ell^{a}\nabla_{a}\ell^{b} = 0 .
\end{equation}
Therefore, there is a natural radial affine function $r$, given by
\begin{equation}
\ell^{a} = \left(\frac{\partial}{\partial r} \right)^{a} .
\end{equation}
Let us note that by construction this congruence is twist free.

In our construction $u=$constant will characterize our past null cone.

The way in which the null function $u$ grows to the future is normalized by the condition
\begin{equation}\label{eq:vconele}
v^a \ell_a = 1;
\end{equation}
where $v^a$ is the observer's 4-velocity. 
This condition implies that we are choosing $\ell^a$ to be future pointing;
and it also fixes the scale of the affine parametrization, 
but it still remains a freedom  associated to the choice of origin; 
or equivalently the 
freedom to make a translation in the value of $r$, for each null direction 
in the congruence.
We will fix below this freedom by taking
the natural choice which correspond to set $r=0$ at the apex of the cone.

In what follows it will be useful to refer the discussion to a null tetrad adapted 
to the past null cone of the observer. 
The first null vector in our tetrad will be $\ell^a$ in the way we have just defined. 
The other vectors are two spacelike vector fields $m^a$ and $\bar{m}^a$ tangents 
to the surfaces $r=$constant and $u=$constant, and satisfying
\begin{equation}
m^a \bar{m}_a = -1;
\end{equation}
and the last null vector will be denoted as $n^a$ which is chosen
orthogonal to $m^a$ and $\bar{m}^a$ and satisfying
\begin{equation}
n^a \ell_a = 1.
\end{equation}

\begin{figure}%[H]
	\centering
	\includegraphics[clip,width=0.4\textwidth]{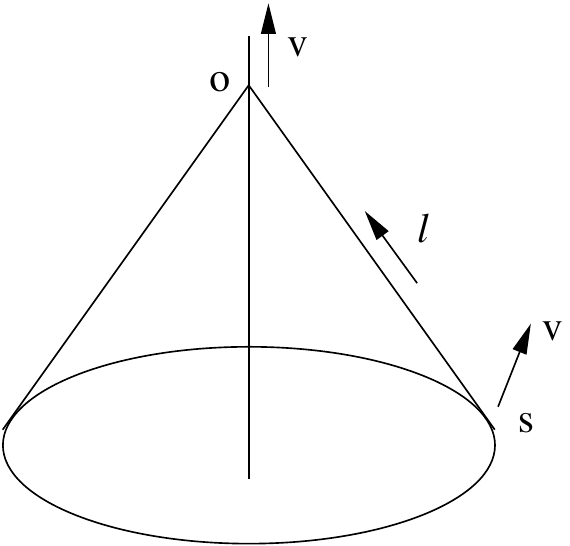}
	\caption{This graph shows schematically a spacetime diagram with
		the location of the observer ``o'', with its velocity,
		a source ``s'', with its velocity, and the trajectory of
		a photon with null vector $\ell$.}\label{fig:lente-cosmo}
\end{figure}

\subsection{The geodesic deviation equation}

The main object for the quantification of the optical distortions is the deviation 
vector $\varsigma^{a}$ which describes the behaviour of the congruence by computing 
the local separation of different geodesics.
By definition it satisfies the condition
\begin{equation}\label{eq: der Lie de varsigma}
\mathscr{L}_{\ell} \; \varsigma^{a} = 0;
\end{equation}
that is, the Lie derivative along $\ell^a$ vanishes.

The equation that provides us with the information about the distortion of images
is the \emph{geodesic deviation equation} for the deviation vector $\varsigma^{a}$ 
along the congruence with tangent $\ell^a$; namely
\begin{equation}
\ell^a \nabla_a \left( \ell^b \nabla_b \varsigma^d \right) = 
R_{abc}^{\;\;\;\;\;d}\ell^a \varsigma^b \ell^c,
\end{equation}
where $R_{abc}^{\;\;\;\;\;d}$ denotes the Riemann tensor.
This equation is obtained by taking a second covariant derivative to the deviation vector.

The deviation vector can be expressed as
\begin{equation}\label{eq. dev vector}
\varsigma^{a} = \varsigma \bar{m}^{a} + \bar{\varsigma} m^{a} ;
\end{equation}
and using the vector $\boldsymbol{\mathcal{X}}$ build up from the components 
$\{ \varsigma, \bar\varsigma \}$;
\begin{equation}
\boldsymbol{\mathcal{X}} = 
\begin{pmatrix}
\varsigma \\
\bar{\varsigma}
\end{pmatrix};
\end{equation}
equation (\ref{eq: der Lie de varsigma}) can be written in the following 
form as
\begin{equation}
\ell(\boldsymbol{\mathcal{X}}) = -P \boldsymbol{\mathcal{X}},
\end{equation}
where the matrix $P$ is
\begin{equation}
P = \begin{pmatrix}
\rho & \sigma \\
\bar\sigma & \bar\rho
\end{pmatrix},
\end{equation}
and $\rho$, $\sigma$ are the spin coefficients in the GHP\cite{Geroch73}
notation which correspond respectively to the expansion and the shear of the 
congruence under study.
We would like to notice that the symbol for the spin coefficient $\rho$
should not be confused with the symbol $\varrho$ that we will later use for
denoting energy density.

It can be seen that the matrix $P$ satisfies
\begin{equation}
\ell(P) = P^2 + Q ;
\end{equation}
where $Q$ is the matrix 
\begin{equation}
Q = 
\begin{pmatrix}
\Phi_{00} & \Psi_{0} \\
\bar{\Psi}_{0} & \Phi_{00}
\end{pmatrix} ;
\end{equation}
whose elements are the curvature components:
\begin{equation}
\Phi_{00} = -\frac{1}{2}R_{ab} \, \ell^{a} \, \ell^{b}  \, ,
\end{equation}
and
\begin{equation}
\Psi_{0} = C_{abcd} \, \ell^{a} \, m^{b} \, \ell^{c} \, m^{d} \, ;
\end{equation}
denoting with $R_{ab}$ the Ricci tensor and $C_{abcd}$ the Weyl tensor.
The equations that describe the behavior of, what we now call the
spin coefficients, have been discussed extensively in the literature,
including the seminal works of R.K. Sachs\cite{Sachs61}.

The second order differential equation coming from the geodesic deviation equation is;
using this notation:
\begin{equation}\label{eq. dev geo}
\ell\left(\ell\left(\boldsymbol{\mathcal{X}}\right)\right) = - Q\boldsymbol{\mathcal{X}}
\; .
\end{equation}
Working with this equation has the advantage that it is explicitly given in terms
of the curvature components.

\section{Null geodesics in Robertson-Walker spacetimes}\label{sec:FRW}
As it is well known R-W metrics are geometries which posses
spatially homogeneous and isotropic symmetry\cite{Wald84} and play 
a fundamental role in our standard description of the Universe at its largest 
scales.

The line element of such geometries can be cast in the following form
\begin{equation}\label{eq:line-element}
ds^2 = dt^2 - A^2(t)dL_k^2;
%\bigg( d\chi^2 + f_{k}^{2}(\chi)d\Sigma^2 \bigg);
\end{equation}
where $t$ is the so-called \emph{cosmological time}, i.e. the proper time 
associated to the family of fundamental observers which perceives isotropy 
and $dL_k^2$ is the line element for the spacelike surfaces orthogonal to 
the preferred observers. 
As a consequence of the symmetries $dL_k^2$ must to be one of the three 
possible 3-Riemaniann metrics of constant curvature. We distinguish them 
by the subindex $k$ which takes the values $k=1$, $k=0$ and $k=-1$  
corresponding respectively to a $3-$sphere, a $3-$plane and a $3-$hyperboloid.
Three equivalent and useful ways of write $dL_k^2$ are:
\begin{align}
dL_k^2 &= d\chi^2 + f_k^2(\chi)d\Sigma^2, \label{eq:dL-standard}
\\ 
dL_k^2 &= \frac{d\mathtt{r}^2}{1 - k\mathtt{r}^2} + \mathtt{r}^2 d\Sigma^2,
\label{eq:dL-standard2}
\\
dL_k^2 &= \frac{1}{\left(1 + \frac{k}{4}\textsf{r}^2 \right)^2}
\left( d\textsf{r}^2 + \textsf{r}^2 d\Sigma^2  \right)
; \label{eq:dL-standard3}
\end{align}
where in equation (\ref{eq:dL-standard}) $f_{k}(\chi)$ is 
given by
\begin{equation}
f_{k}(\chi) = 
\begin{cases}
\sinh(\chi), & \text{for  $k=-1$}, \quad \text{$0 \leqslant \chi < \infty$}, 
\\
\chi, & \text{for $k=0$}, \quad \text{ $0 \leqslant \chi < \infty$}, 
\\
\sin(\chi), & \text{for $k=1$}, \quad \text{$0 \leqslant \chi \leqslant \pi$}
.
\end{cases}
\end{equation}
The two-dimensional line element $d\Sigma^2$ is the common metric of the
unit $2-$sphere, which is often found in one of the two following forms
\begin{align}
d\Sigma^2 &=d\theta^2 + \sin^2 (\theta) d\phi^2, \\
d\Sigma^2 &= \frac{4}{\left( 1 + \zeta\bar{\zeta}\right)^2}d\zeta d\bar{\zeta} 
\equiv \frac{1}{P_0^2}d\zeta d\bar{\zeta};
\label{eq:ds-esfera-complex}
\end{align}
where $\left( \theta, \phi \right)$ are usual angular coordinates and
$\left( \zeta, \bar{\zeta} \right)$ complex stereographic coordinates 
on the complex plane.

\subsection{Past light cone of preferred observers and associated null tetrads}
Regarding to the information inferred from observations in our past light cone 
one realizes that the relevant geometrical quantities for the description of 
observations are the null rays reaching the observer and the angular deviations 
of such rays. All observable quantities are constructed from them.
Then, it is necessary to have at hand an adequate description in terms of the 
past light cone of the observers. 

Associated to the observer one has its proper time $\tau$; which for 
co-moving observers coincides with the standard coordinate time $t$. 
Then, the $4-$velocity $v^a$ of the observer, can be expressed as
\begin{equation}
v^a = \left(\frac{\partial}{\partial t}\right)^a.
\end{equation}

In the following we will introduce a coordinate system which is well adapted to 
the discussion that follows.
We begin by labeling each past light cone with a null function $u$ which takes
constant values on the cones. In particular, this means that 
the value of $u$ can be set by the proper time $\tau$; 
an we choose the relation
\begin{equation}
\frac{du}{d\tau} = 1 ;
\end{equation} 
at the origin;
but one can also use another specification for $u$ that is not based on
the synchronization of the null function with the proper time of the observer.
In fact, we will use below a definition based on the conformal structure
of the metric. 
But in any case, the physically relevant null geodesic congruence
associated to the observer, can be expressed in terms of this null function
by (\ref{eq:ell_du}).

Let us note also that this setup can be generalized 
for an arbitrary world line $\gamma'_0$ with $4-$velocity $v'^a$ in terms of a 
general proper time $\tau'$.

\subsubsection{The conformal advanced null function $\tilde{u}$}\label{subsec:conformal-u}
Let us introduce another null function which is very useful for the calculations. 
We define $\tilde u$ from
\begin{equation}\label{eq:du}
d\tilde u = \frac{dt}{A(t)} + d\chi;
\end{equation}
where we have used tilde in order to differentiate it from the natural 
choice presented before.

This choice is appropriated to the study of the asymptotic behaviour 
of the geometry \cite{Moreschi:1988ik} but it is also helpful in many computation 
in comparison with the choice $u = \tau$.

The next step is to complete the coordinate system putting coordinates on each 
past light cone. 
In this process also a null tetrad $(\tilde\ell^a, \tilde m^a, \tilde{\bar{m}}^a, 
\tilde n^a)$ associated to this construction can be selected.

First let us consider the vector $\tilde\ell^a$ determined by the one-form
\begin{equation}
\tilde\ell = d\tilde u;
\end{equation}
which is geodesic and affinely parametrized, since $\tilde u$ is null.

Now, let $\tilde r$ be the affine parameter along the cone associated to 
$\tilde\ell^a$, starting from the observer. 
This will be our second coordinate. 
Then, a pair of null complex vectors $\tilde m^a$ and $\tilde{\bar{m}}^a$ can be 
chosen to be tangent to the surfaces $\tilde u=$constant and $\tilde r=$constant 
and satisfying
\begin{equation}
\tilde m^a \tilde{\bar{m}}_a = -1.
\end{equation}

To complete our coordinate system we label the generators of the null cone by 
the stereographic coordinates $\left( \zeta, \bar{\zeta}\right)$.
Finally, we also complete our tetrad with a null vector 
$\tilde n^a$ chosen such that 
\begin{align}
\tilde n^a \tilde m_a &= 0, \\
\tilde n^a \tilde\ell_a &= 1.
\end{align}

Explicitly, in term of the usual coordinates $(t,\chi, \zeta, \bar{\zeta})$, 
the vector field $\tilde\ell^a$ is expressed as:
\begin{equation}
\tilde\ell = \frac{1}{A(t)}\left( \frac{\partial}{\partial t}\right) - 
\frac{1}{A^{2}(t)} \left( \frac{\partial}{\partial \chi}\right) ;
\end{equation}
while for the other tetrad vectors we have:
\begin{align}
\tilde m & 
= \frac{\sqrt{2}P_0}{A(t) f_k(\chi)}\left(\frac{\partial}{\partial \zeta} \right) , 
\\
\tilde{\bar{m}} & 
= \frac{\sqrt{2}P_0}{A(t) f_k(\chi)}\left(\frac{\partial}{\partial\bar{\zeta}}\right)
, \\
\tilde n &= \frac{A(t)}{2}\left[\left(\frac{\partial}{\partial t} \right) 
+ \frac{1}{A(t)}\left(\frac{\partial}{\partial \chi} \right) \right].
\end{align}

The line element  (\ref{eq:line-element}) with the choice 
(\ref{eq:dL-standard}) can be expressed in terms of the new coordinate system 
$\left(\tilde u, \tilde r,\zeta,\bar{\zeta} \right)$.

We will do this with an intermediate step; using eq. (\ref{eq:du}) one can express
the geometry in terms of the coordinate system $(\tilde u,\chi,\zeta,\bar\zeta)$
to find:
\begin{equation}
ds^2 = A^2(t) \left(
d\tilde u^2 - 2 d\tilde u d\chi - f_k^2 d\Sigma^2 
\right) ;
\end{equation}
where here it is important to note that the vector $\tilde{\ell}$ is expressed as
\begin{equation}\label{eq: ell geodesico}
\tilde \ell = - \frac{1}{A^2\left(t(\tilde u,\chi)\right)}
\left(\frac{\partial}{\partial \chi} \right).
\end{equation}
Since also $\tilde \ell = \frac{\partial}{\partial \tilde r}$, we also 
have the relation
\begin{equation}
\frac{\partial\chi}{\partial\tilde{r}} = - \frac{1}{A^2 \left(t(\tilde u,\chi)\right)};
\end{equation}
so we take
\begin{equation}
\tilde{r} = - \int_{0}^{\chi} \,A^2\left(t(\tilde u,\chi')\right) d\chi' = 
\int_{t(\tilde u,\chi=0)}^{t(\tilde u,\chi)} \,A(t')dt'.
\end{equation}

This is the relation that we need to cast the line element in terms
of the coordinates $\left(\tilde u, \tilde r,\zeta,\bar{\zeta} \right)$
since now we can differentiate to obtain	
\begin{equation}
d\tilde r = 
\bigg(A^2 \big(t(\tilde u,\chi) \big) - A^2(t_0) \bigg)d\tilde u - 
A^2\big( t(\tilde u,\chi) \big)d\chi;
\end{equation}
where we use the notation $A(t_0)$ to mean
\begin{equation}
A(t_0) \equiv A\big(t(\tilde u,\chi=0) \big); 
\end{equation}
where it should be noted that $t_0 = t(\tilde u,\chi=0)$ need not be the present time;
but the coordinate $t$ evaluated at the origin.

Then, one arrives to the desired expression for the metric:
\begin{equation}
ds^2 = \bigg(2 A^2(t_0) - A^2(t) \bigg)d\tilde u^2 + 2d\tilde u 
d\tilde r - A^2(t) f_k^2(\chi)d\Sigma^2.
\end{equation}

For completeness we include here the relations among the differential of the old
coordinate system in terms of the latest one; they are given as follows:
\begin{align}
dt =& \frac{A^2(t_0)}{A(t)} d\tilde u + \frac{1}{A(t)} d \tilde r,
\\
d\chi =& \bigg(1 - \frac{A^2(t_0)}{A^2(t)} \bigg) d\tilde u - 
\frac{1}{A^2(t)}  d\tilde r .
\end{align}

\subsubsection{The inertial advanced null function $u$}\label{subsec:inertial-u}
In this case we consider the choice $u = \tau$ which will be presented 
in similar way to the discussion of the previous section. 
This has the advantage to help the reading and at the same time it allows 
to arrive to the main expression in a simple way using the definitions
presented above.

Then, let us find the relation between both functions by noting that
\begin{equation}
v \left( \tilde{u} \right) = \frac{d\tilde{u}}{du};
\end{equation}
along the world line of a preferred observer. 
Since $u$ is normalized with $t$ at the origin, from equation 
(\ref{eq:du}) one has
\begin{equation}\label{eq: du-tilde}
du 
= A(t_0) d\tilde u 
= A\big(t(\tilde u,\chi=0) \big) d\tilde u
;
\end{equation}
which leads us to define the first null vector $\ell$ as
\begin{equation}
\ell = d u .
\end{equation}
This gives us a congruence of future directed null geodesics, affinely 
parametrized that reach the observer. The affine parameter in this case 
will be denoted by $r$ and constitutes our second coordinate. 
Stereographic coordinates label the generators of the congruence 
as before.

The null vector $\ell$ can be expressed in the following way:
\begin{equation}
\ell = \frac{A(t_0)}{A(t)}\left[ \left(\frac{\partial}{\partial t} 
\right) - \frac{1}{A(t)}\left(\frac{\partial}{\partial \chi} \right) \right];
\end{equation}
while the other vectors of the tetrad can be taken to be
\begin{align}
{m} &=  \frac{\sqrt{2}P_0}{A(t) f_k(\chi)}\left(\frac{\partial}{\partial \zeta} \right),
\\
{\bar{m}} &=  \frac{\sqrt{2}P_0}{A(t) f_k(\chi)}
\left(\frac{\partial}{\partial \bar{\zeta}} \right), 
\\
{n} &= \frac{1}{2}\frac{A(t)}{A(t_0)}\left[\left(\frac{\partial}
{\partial t} \right) + \frac{1}{A(t)}\left(\frac{\partial}{\partial \chi} \right) \right]
.
\end{align}

We left for the appendix \ref{ap:Geom-Null} a discussion with the 
details of the connection and the curvature of the geometry using
the null tetrad introduced above.

The metric in the advanced coordinates $\left( u, \chi, \zeta, \bar{\zeta} \right)$
can be expressed as
\begin{equation}
ds^2 = A^2(t) \bigg( \frac{du^2}{A^2(t_0)} - 2\frac{d u d\chi}{A(t_0)} 
- f_k^2(\chi)d\Sigma^2 \bigg).
\end{equation}

The affine parameter $r$ associated to ${\ell}$ and the differential $dr$ are found 
to be 
\begin{equation}\label{eq: r tilde}
{r}( u, \chi) 
= \frac{1}{A(t_0)} \int_{t( u, \chi=0)}^{t( u, \chi)}{A(t')dt'},
%= \frac{\tilde r}{A(u,\chi=0)},
\end{equation}
and
\begin{equation}\label{eq:dr-dchi-du}
dr = -\frac{A^2( u, \chi)}{A(t_0)} d\chi 
- \left( 1 - \frac{A^2( u,\chi)}{A^2(t_0)} + r H(t_0) \right)du ;
\end{equation} 
where we introduce here the \emph{Hubble function} on the observer world
line 
\begin{equation}
\begin{split}
H(t_0) \equiv 
\frac{1}{A(t_0)}\frac{dA(t({u}, \chi= 0))}{d{u}}.
\end{split}
\end{equation}

One can then transform to a coordinate system in which the radial coordinate
is $r$ instead of $\chi$; so that now $t=t(u,r)$ and $\chi(u,r)$.
Note that now $t_0=t(u,r=0)$.
Gathering together this expressions one arrives to the metric in coordinates
$\left( u, r, \zeta, \bar{\zeta} \right)$:
\begin{equation}
\begin{split}
ds^2 =& 2
\left(1 -\frac{A^2(t)}{2 A^2(t_0)} + r
H(t_0) \right) d u^2 + 2d u d{r} \\
& - A^2(t)f_k^2(\chi)d\Sigma^2 .
\end{split}
\end{equation}

\subsection{Deviation vector for a radial congruence of null geodesics}
In the R-W geometries, the behaviour of the deviation vector, is easily found 
by considering the ordinary linear differential equation (\ref{eq: der Lie de varsigma})	
on the past null cone:
\begin{equation}\label{eq:l-rho=-rho-chi}
\ell(\boldsymbol{\mathcal{X}}) = - \rho \boldsymbol{\mathcal{X}}.
\end{equation}

Let us note here that we can use $\ell^a$ or $\tilde{\ell}^a$ as vector field
tangent to the congruence since the description is independent of the affine 
parametrization.

In the coordinates $\left( u, r \right)$ of section \ref{subsec:inertial-u}
the convergence $\rho$ takes the simple and transparent expression:
\begin{equation}\label{eq:rho-Area-RW}
\rho = - \frac{1}{A(t)f_k(\chi)}\ell\left( A(t) f_k(\chi) \right).
\end{equation}
This means that the deviation vector is given by:
\begin{equation}\label{eq: X solution}
\boldsymbol{\mathcal{X}} = A(t) f_k(\chi) \boldsymbol{\mathcal{V}}_0;
\end{equation}
where the integration constant $\boldsymbol{\mathcal{V}}_0$ has the meaning
of the angle measured by the fundamental observer between two neighbour 
null geodesic.

Since this factor will appear frequently we define
\begin{equation}\label{eq:D-A-fisrt-defin}
D_A(t,\chi) \equiv A(t) f_k(\chi);
\end{equation}
which will be identified below to the angular diameter distance.

\section{Different notions of distance}\label{sec:distances}

Since the notion of distance plays a central role in the discussion of gravitational lenses,
and it is fundamental to all the discussions in this paper, 
it is convenient that we
review here the usual notions of distance that appear in the cosmological
framework.

\subsection{The redshift}
The observer $o$ defines the redshift associated with the event $s$ by the expression
\begin{equation}\label{eq:redshift}
1 + z = \frac{\left(l^a v_a\right)_{s}}{\left(l^a v_a\right)_o};
\end{equation}
where $l^a$ is the tangent vector to a null geodesic connecting the two events 
and $v^a$ denotes either the $4-$velocities of the source $s$ or that of the observer. 

Although the redshift is an indication of different proper times measures it is also
used as an indication of distance in R-W spacetimes.
When the source and the observer are both fundamental observers of a R-W spacetime, one has
\begin{equation}
1 + z = \frac{A(t_0)}{A(t)}.
\end{equation}
In other words, the redshift can be used as a notion of distance if one has at hand
a given family of cosmological objects with known motion.

\subsection{Geometric distance or normalized affine distance}\label{subsec:Geom-distance}
Given an observer with  4-velocity $v^a$, there is a natural notion of distance $\lambda$ 
defined on any past directed null geodesic on any spacetime; which is identified with the 
affine parameter so chosen that normalization (\ref{eq:vconele}) holds, 
and is zero at the observer position. 
In the case of a R-W spacetime, this distance is related to the coordinate and affine 
parameter $r$, by
\begin{equation}
\lambda = - r .
\end{equation}
To differentiate from other definitions we will call this the \emph{geometric distance}
or \emph{normalized affine distance}.
By definition it increases monotonically along every past directed null geodesic
from the observer.

A fundamental concept in the study of gravitational lenses is the comparison
of an image that is observed, with the unaffected image, for which gravitational
effects can be completely neglected.
But, then, it is crucial to compare the images in both situations when the
source is at the `same' distance.
It is here that the definition of the distance that one should use, becomes
important, and we claim that the appropriate notion is the one given
by the definition of normalized affine distance, or geometric distance.
This is the concept 
that appears as the dynamical parameter in the fundamental
equation  (\ref{eq. dev geo}),
that is always well behaved (it increases monotonically
as the objects are further away) and it 
can be applied to any situation and spacetime
(with or without symmetries), and that we use as the fundamental
notion of distance in this work.

\subsection{Angular diameter distance and future angular distance}
\subsubsection{Angular diameter distance and area distance}

The so called `angular diameter distance', $D_A$, is defined as
\begin{equation}
dl = D_A d\theta;
\end{equation}
where $dl$ is the projected size of the object, perpendicular to the line of sight, 
at the place of the emitter,
%by a bundle of null geodesics diverging from the observer which subtends a solid 
and $d\theta$ is the subtended angle measured in the sky by the observer.

This definition is closely related to the \emph{area distance};
which is defined from:
\begin{equation}
dA = |D_+ D_-| d\Omega;
\end{equation}
where $dA$ is the projected area of the object, perpendicular to the line of sight, 
at the place of the emitter, $D_+$ and $D_-$ are the angular diameter distances
along the principal directions of the area\cite{Perlick:2004tq},
and $d\Omega$ is the solid angle seeing by the observer;
and the area distance $D_\text{area}$ is defined by
\begin{equation}
D_\text{area} = \sqrt{ |D_+ D_-|} ;
\end{equation}
so that now one can write
\begin{equation}\label{eq:dAdarea-dOmega}
dA = D_\text{area}^2 \, d\Omega .
\end{equation}

Let us note that this measure of distance does not depend on the peculiar velocity of 
the emitter, it depends only on the $4-$velocity of the observer\cite{Perlick:2004tq}.

It is probably important to emphasize that in the discussion on R-W spacetimes the 
area distance coincides with the angular diameter distance; for this reason in the 
remainder of the paper will just use $D_A(=D_\text{area})$ to denote both distances,
in the case of R-W spacetimes.

For fundamental observers in R-W geometries this notion of distance is explicitly
given by those presented in equation (\ref{eq:D-A-fisrt-defin}):
\begin{equation}\label{eq:da}
D_A = A\left(t({u}, {r}) \right)
f_k\left(\chi({u}, {r}) \right).
\end{equation}
From its definition, one notes that the \emph{area distance} is related
to the expansion of a thin bundle of null geodesics leaving the observer and reaching
the source; in fact, in a R-W spacetime one can see this fact explicitly by looking
at equation (\ref{eq:rho-Area-RW}). In this particular case one has
\begin{equation}
\frac{dD_A}{dr} = - \rho D_A;
\end{equation}
and the second order differential equation
\begin{equation}
\frac{d^2D_A}{dr^2} = -\Phi_{00}D_A.
\end{equation}
It can be seen that 
in general $\rho$ has a value $r^*$ for which it vanishes and then change its sign.
Then, the value of $D_A$ reaches an stationary value at $r^*$; which is a maximum,
since $\Phi_{00}$ is positive when the usual energy conditions are considered to hold.

For this reason this notion of distance is not really a distance;
since in particular it tends to zero when the source is at the farthest distance,
i.e. at the beginning of the Universe; where $A \rightarrow 0$.
But in any case it is a useful concept that can be employed to describe the behaviour of 
gravitational lenses.

\subsubsection{Future angular diameter distance and future area distance}
Let us also note that there is a notion of \emph{future angular diameter distance} $d_{A}$, 
which is related to the angular diameter distance calculated by the emitter to reach 
an object of projected size $dl_o$ at the place of observer, perpendicular to the direction 
of the path of a photon.
More concretely, one defines $d_{A}$ by the relation
\begin{equation}
dl_o = d_{A} \, d\theta_s;
\end{equation}
where 
$d\theta_s$ is the angle measured by a fiducial observer at 
the source with the same $4-$velocity as the source.

We can also generalize to the notion of \emph{future area distance};
which is defined from:
\begin{equation}\label{eq:dadomega}
dA_o = |d_+ d_-| d\Omega_s;
\end{equation}
where $dA_o$ is the projected area of the object at the place of the observer,
and $d\Omega_s$ is the solid angle generated at the source position,
$d_+$ and $d_-$ are the future angular diameter distances
along the principal directions of the area;
and the future area distance $d_\text{area}$ is defined by
\begin{equation}
d_\text{area} = \sqrt{ |d_+ d_-|} .
\end{equation}

Also in the case of a R-W geometry the future area distance coincides with the 
future angular diameter distance and 
so one can use $d_{A}(=d_\text{area})$ for both notions.

We will discuss in section \ref{subsec:reciprocity-theorem} the existing relation between
the future area distance $d_{A}$ and the area distance $D_A$.
For the moment, we will mention that in the most general case of an \emph{unresolved}
non-isotropic radiating  source at redshift $z$ which emits with a 
luminosity $L(\theta_s, \phi_s)$ in the direction $(\theta_s, \phi_s)$, pointing out into the
observer direction, one predicts that the flux $\mathscr{F}(\theta_o, \phi_o)$ measured by the 
observer is given by 
\begin{equation}\label{eq:flux-lumin}
\mathscr{F}(\theta_o, \phi_o) \, dA_o = \frac{1}{( 1 + z )^2} L(\theta_s, \phi_s) \, d\Omega_s ;
\end{equation}
where the left hand side is the expression for the energy measured by the 
observer per unit time in the direction of the source $(\theta_o, \phi_o)$,
collected in the surface element $dA_o$, 
and 
the right hand side is the expression for the energy emitted by the source
per unit (local)time in the direction of the observer, in the 
solid angle $d\Omega_s$, with the corresponding redshift factor that
takes into account the difference in local proper times, and the 
local energy measures. 
To clarify the notation, let us remark that if we
call $L_0$ the total luminosity, and if the source where radiating
isotropically, then one would have $L = L_0/(4 \pi)$.

Let us note that another way to write (\ref{eq:dadomega}) is:
\begin{equation}\label{eq:dadomega-2}
dA_o = d_\text{area}^2 \, d\Omega_s :
\end{equation}
so that one finally has
\begin{equation}\label{eq:flux-future-ang-distance}
\mathscr{F}(\lambda,z) = 
\frac{L(\theta_s, \phi_s)}{\left(1 + z \right)^2 d_{\text{area}}^2(\lambda)};
\end{equation}
where we are emphasizing that the flux depends on the distance $\lambda$ and on
the state of motion of the source, characterized by the redshift factor of 
expression (\ref{eq:redshift}).

Let us note that (\ref{eq:flux-future-ang-distance}) is valid in a general spacetime
and that what we call future angular diameter distance is what other authors
call `corrected luminosity distance'\cite{Perlick:2004tq}: but we prefer our wording
because it is more natural for the discussion of the \emph{reciprocity theorem}
that we present below.

\subsection{The Etherington theorem and the reciprocity corollary}\label{subsec:reciprocity-theorem}
It is worthwhile to mention that the definition of area distance $D_\text{area}$
and future area distance $d_\text{area}$ have been presented for a general spacetime.
For the general case there is a purely geometric result due to 
Etherington\cite{Etherington33} which establishes that\cite{Ellis71}:
\begin{equation}
dA_o \, d\Omega = \left( 1 + z \right)^2 dA \, d\Omega_s ;
\end{equation}
which is known as the \emph{Etherington theorem}.
An interesting corollary of this result is derived from
(\ref{eq:dAdarea-dOmega}) and (\ref{eq:dadomega-2}), since
one can prove the relation between the area distances, namely
\begin{equation}\label{eq:reciprocityth}
d_{\text{area}} = \left( 1 + z \right) D_\text{area} .
\end{equation}

This is a fundamental relation between the future area distance $d_{\text{area}}$ 
and the (observed) area distance $D_\text{area}$,
also known as the \emph{reciprocity relation}, which 
is only deduced from considerations of thin bundles connecting the
source and the observer.

Also, relation (\ref{eq:reciprocityth}) is the main tool to prove the 
distance-duality relation, that we recall next.

\subsection{Luminosity distance and the distance duality-relation}
Let us return to the relation between the observed flux and the total luminosity
of an unresolved emitter, shown in equation (\ref{eq:flux-future-ang-distance}). 
This relation gives rise to the notion of \emph{luminosity distance} $D_L$
which is defined
in term of $\mathscr{F}(\theta_o, \phi_o)$ and $L(\theta_s, \phi_s)$ in a completely analogous way 
than in a flat-spacetime, namely 
\begin{equation}
D_L \equiv \sqrt{\frac{L(\theta_s, \phi_s)}{\mathscr{F}(\theta_o, \phi_o)}};
\end{equation}
from which one can see that:
\begin{equation}\label{eq:lum-dist-vs-fut-ang-dist}
D_L = \left( 1 + z \right) d_{\text{area}}.
\end{equation}

Although this definition has been presented for unresolved sources;
estimates of distances from extended sources 
can be based on the same principle. 
In such cases the problem to infer distance for \emph{resolved} or 
\emph{extended objects} 
involves the consideration of the so called \emph{surface brightness}.
We will discuss these topics elsewhere.

\subsubsection{Distance-duality relation}
The combination of equation (\ref{eq:lum-dist-vs-fut-ang-dist}) together with the 
geometric relation established by the reciprocity 
theorem (\ref{eq:reciprocityth}) gives as a corollary, the
so called \emph{distance-duality relation}:
\begin{equation}
D_L(z) = \left( 1 + z \right)^2 D_\text{area};
\end{equation}
which links the luminosity distance with the area distance.

It is worthwhile to remark that this relation,
valid for a general spacetime,
has been derived just from 
a couple of assumptions, namely:
that photons follow geodesics and that the number of photons is conserved.
Since, the geodesic motion of photons is sustained by the undisputed
equivalence principle,
any observed deviations from the distance-duality relation would imply a 
non-conservation of the mean number of photon along null 
geodesics\cite{Bassett:2003vu}.

\section{Different notions of magnifications for general spacetimes}\label{sec:magnifications}

\subsection{Angular magnification}

In a general spacetime the \emph{angular magnification} $\mu$,
which is normally addressed just as \emph{the magnification},
is defined as the 
ratio between the solid angles subtended by the ``lensed'' and the ``unlensed'' image.
When using the principal directions of the lensed image, one
has the relations
\begin{equation}\label{eq:betacosmic--}
\delta \beta_- = \left( 1 - \kappa - \gamma \right)  \delta \theta_- ,
\end{equation}
and
\begin{equation}\label{eq:betacosmic-+}
\delta \beta_+ = \left( 1 - \kappa + \gamma \right)  \delta \theta_+ ;
\end{equation}
from which one can also deduce that
\begin{equation}
\label{eq:mu-2}
\mu = \frac{1}{(1-\kappa)^2 - \gamma^2 }
;
\end{equation}
where $\gamma^2 = \gamma_1^2 + \gamma_2^2$ is the square of the module of the shear.

For most astrophysical situations one has $\kappa << 1$ and $\gamma << 1$.
If we multiply, let us say, (\ref{eq:betacosmic--}) by $\lambda$ we obtain
\begin{equation}\label{eq:dl-}
dl_- = \lambda \delta \beta_- = \lambda\left( 1 - \kappa - \gamma \right)  \delta \theta_- 
= D_- \delta \theta_-
,
\end{equation}
and similarly with the + case.
Then it is deduced that one can also express
\begin{equation}
\label{eq:mu}
\mu = \left(\frac{\lambda}{D_\text{area}(\lambda)}\right)^2 ;
\end{equation}
since the area distance in flat spacetime is just $\lambda$.
This expression, valid for a general spacetime, 
coincides with the definition found in eq. (42) of the reference 
\cite{Perlick:2004tq}.

\subsection{The intensity magnifications}

In this subsection we present the natural physical intensity magnification,
that we will denote with $\tilde{\mu}$, and that we just call \emph{intensity magnification};
but also it is  convenient to introduce the  astrophysical motivated
\emph{cosmic intensity magnification},
that we will denote with $\mu'_c(z)$.
The behavior of these intensity magnifications will be discussed
below after the introduction of a couple of cosmic examples.

Let us denote again with $\mathscr{F}$ the observed flux of an 
unresolved object at distance $\lambda$ 
and relative motion determined by the redshift $z$,
in a general spacetime.
We will use $\mathscr{F}_0$ to denote the flux
that one expects to collect from the same object at the same distance $\lambda$ 
in Minkowski spacetime with the same relative motion.
Then, we define the intensity magnification $\tilde{\mu}$ 
by the quotient of these two fluxes, namely:
\begin{defi} Intensity magnification:
	\begin{equation}\label{eq:mag_lambda}
	\tilde{\mu}(\lambda) \equiv \frac{\mathscr{F}(\lambda,z)}{\mathscr{F}_0(\lambda,z)} 
	= \bigg( \frac{d_\text{area0}(\lambda)}{d_\text{area}(\lambda)} \bigg)^2
	;
	\end{equation}
	where, taking into account equation (\ref{eq:flux-future-ang-distance}),
	one can see that actually $\tilde\mu$ turns out to depend only on 
	the distance $\lambda$, and we use the subindex 0 to denote quantities
	in Minkowski spacetime.
\end{defi}

It is probably worthwhile to remark that we are using here
the same philosophy that one applies to the basic notion of gravitational
lens in equation (\ref{eq:standarlensequation}), where one compares angles
in the sphere of directions between the lensed situation and the 
unlensed one (flat case).

It is crucial here to recall the reciprocity theorem, since it allows us
to prove the following theorem, valid for a general spacetime:
\begin{teo}\label{teo:rema}
	
	The intensity magnification coincides with the (standard angular) magnification;
	namely:
	\begin{equation}
	\tilde \mu = \mu .
	\end{equation}
	
\end{teo}
Proof: We have seen above that a corollary of the Etherington theorem, presented in
\ref{subsec:reciprocity-theorem}, is the reciprocity relation (\ref{eq:reciprocityth});
so that in the expression for the intensity magnification above 
one can replace the future area distances by the 
observed area distances, and express 
\begin{equation}
\begin{split}
\tilde{\mu}(\lambda) = 
\left(\frac{D_{\text{area}0}(\lambda) }{D_\text{area}(\lambda)} \right)^2 
=
\left( \frac{\lambda}{D_\text{area}(\lambda)} \right)^2 
=
\mu
;
\end{split}
\end{equation}
where, we have used equation (\ref{eq:mu}).

%\vspace{2mm}
It should be noted that this is a relation valid in a general spacetime
which relates the here defined observable \emph{intensity magnification},
with the standard angular magnification;
instead the usual discussion is done relating
the standard magnification with the non-observable luminosity distance 
as applied in spherically symmetric spacetimes.
This is normally carried out in Robertson-Walker geometries that
are spherically symmetric around every point of the spacetime.
If the exact geometry of the Universe where represented by a Robertson-Walker geometries
then it would be a matter of personal choice, whether one would like
to use the concept of luminosity distance or of the observed flux;
but we do know that the Universe is not homogeneous nor isotropic,
and therefore it is important that theorem \ref{teo:rema} is true
for general spacetimes.

A very important point to remark is that astrophysicists normally think in the 
relation of the fluxes as a function of the observed redshift; so that
from the observational point of view, it might become useful also to define
\begin{defi} Cosmic Intensity Magnification:
	\begin{equation}\label{eq:mag_z}
	\begin{split}
	\mu'_c(z) \equiv& 
	\frac{\mathscr{F}(z)}{\mathscr{F}_{\text{Milne}}(z)} 
	= \left( \frac{ 1 + z }{ 1 + z} \right)^4 \left(\frac{D_{A0}(z) }{D_A(z)} \right)^2 \\
	=&  \left(\frac{\lambda_\text{Milne}(z) }{D_A(z)} \right)^2 
	;
	\end{split}
	\end{equation}
	where more specifically in this case one uses the flat Milne cosmological
	model, in order to have a relation $\lambda(z)$.
\end{defi}

Here it should be remarked that although the definition of this intensity
magnification also involves the quotient of the same fluxes;
the definition (\ref{eq:mag_lambda}) gives a different function
because it involves a different parametrization of the scalars.
For this same reason, the right hand side of (\ref{eq:mag_z}) will not 
coincide with the corresponding cosmic magnification $\mu_c$ 
that it will be defined below in equation (\ref{eq:mu-cosm}).
In particular, since in the flat spacetime case we need a relation between
the distance $\lambda$ and the redshift $z$, one makes use of the cosmological
model given by the flat Milne Universe, that is recalled in the next subsection.

The reason we introduce the intensity magnification $\tilde \mu$ is 
because it is
directly related to observations; since astrophysicists measure relative
apparent magnitudes, as given by\cite{bradt2004astronomy} 
\begin{equation}
m_2 - m_1 = -\frac{5}{2} \log \frac{\mathscr{F}_2}{\mathscr{F}_1}
.
\end{equation}
The collecting apparatus normally does not receive all the photons but those
on a limited band; we assume here for the sake of simplicity in the discussion
that all photons are collected; but one can generalize these concepts to the realistic
situation.

To obtain intrinsic information, one can also use the so called \emph{absolute magnitude} $M$;
which is defined as the magnitude of the object if it were at the standard distance of 10
parsecs.
The relation of the apparent magnitude and the absolute magnitude is given by
\begin{equation}\label{eq:m-M}
\begin{split}
m - M 
=& -\frac{5}{2} \log \frac{\mathscr{F}(\lambda,z)}{\mathscr{F}(\lambda_{10},z_{10})} 
=  5 \log \frac{D_L(\lambda,z)}{D_L(\lambda_{10}(z_{10}),z_{10})}
\\
=& -\frac{5}{2} \log 
\bigg( 
\frac{\mathscr{F}(\lambda,z)}{\mathscr{F}_{0}(\lambda,z)} 
\frac{\mathscr{F}_0(\lambda,z)}{\mathscr{F}_{0}(\lambda_{10},z_{10})}
\frac{\mathscr{F}_{0}(\lambda_{10},z_{10})}{\mathscr{F}(\lambda_{10},z_{10})}
\bigg) \\
=& -\frac{5}{2} \log \tilde\mu(\lambda)
+ 5 \log
\bigg( 
\frac{\lambda (1+z)^{2} }{\lambda_{10}}
\bigg)
;
\end{split}
\end{equation}
where the first term in the last line corresponds to the first factor
in the logarithm, the second term corresponds to the second factor
in the logarithm, and 
we have neglected the contribution coming from the ratio
$\frac{\mathscr{F}_{0}(\lambda_{10})}{\mathscr{F}(\lambda_{10})}$;
since the cosmic incidence to the curvature for distances up to 10pc 
can be completely ignored.

We have used here the notation that $\mathscr{F}$ refers to the flux measured 
in the real (general) spacetime 
and $\mathscr{F}_{0}$ refers to the flux measured in the situation of
empty (flat) spacetime.

It is important to emphasize that the left hand side of
equation (\ref{eq:m-M}) is known as \emph{distance modulus},
and that it can be completely expressed in terms of the
gravitational lens optical scalars, since due to theorem \ref{teo:rema}
one can write
\begin{equation}\label{eq:m-M_2}
\begin{split}
m - M 
= -\frac{5}{2} \log \mu(\lambda)
+ 5 \log
\bigg( 
\frac{\lambda (1+z)^{2} }{\lambda_{10}}
\bigg)
;
\end{split}
\end{equation}
where the magnification $\mu(\lambda)$ can be understood in terms of
(\ref{eq:mu-2}) or (\ref{eq:mu}); which are valid for general spacetimes.
Let us also note that the second term in the right hand side of 
(\ref{eq:m-M_2}) has kinematical information, and has no dependence
on any possible cosmological background.

Let us now express the relation between magnifications when the redshift is used
as the single variable to denote distance, as is normally done in cosmological studies.
The previous expression involving the absolute magnification is now given by:
\begin{equation}\label{eq:m-M_astro}
\begin{split}
m - M 
=& -\frac{5}{2} \log \frac{\mathscr{F}(\lambda(z),z)}{\mathscr{F}(\lambda_{10}(z_{10}),z_{10})} \\
=& \;  5 \log \frac{D_L(\lambda(z),z)}{D_L(\lambda_{10}(z_{10}),z_{10})}
\\
=& -\frac{5}{2} \log 
\bigg( 
\frac{\mathscr{F}(\lambda(z),z)}{\mathscr{F}_{\text{Milne}}(\lambda(z),z)} 
\frac{\mathscr{F}_{\text{Milne}}(\lambda(z),z)}{\mathscr{F}_{\text{Milne}}(\lambda_{10}(z_{10}),z_{10})} \\
&\frac{\mathscr{F}_{\text{Milne}}(\lambda_{10}(z_{10}),z_{10})}{\mathscr{F}(\lambda_{10}(z_{10}),z_{10})}
\bigg) \\
=& -\frac{5}{2} \log \mu_c'(z)
+ 5 \log
\bigg( 
\frac{\lambda_{\text{Milne}}(z) \; (1+z)^{2} }{\lambda_{10}}
\bigg)
;
\end{split}
\end{equation}
where again we have neglected the cosmic effects at the short distance of 10pc.
One can see that in this case the expression is given in terms of the
intensity magnification, $\mu'_c(z)$.

\section{Robertson-Walker spacetime as a gravitational lens}\label{sec:RW-Lens}

\subsection{The fundamental concepts in the cosmological context}
It is important to remark that the basic notion of a gravitational lens
is encoded in equation (\ref{eq:standarlensequation}), in which the
variation angle $\delta \beta^a$ refers to an unaffected spacetime,
and therefore Minkowski; while the variation angle $\delta\theta^b$
refers to the real spacetime, in this case R-W.
Then, 
when
one considers this spacetime as a gravitational lens, this is not located 
at a particular position, it rather fills the whole spacetime.
It is because of this reason that one has to be very careful with
the interpretation of the optical scalars, as defined in equation (\ref{eq:optical-scalars0}).

We are thinking in the case of an observer within a R-W spacetime which,
when studying the variations of deviation angles, 
he must relate the observed angle $\delta \theta^a$
with   angle $\delta \beta^a$, when the source is
at the `same' \emph{distance};
where as usual, the second angle refers to an observation where no gravitational
effects are present.
Therefore it is crucial to have a clear understanding of the notion of
\emph{distance}, that one must use.
From all the notions of distance that are employed in the literature and that
we have mentioned in section \ref{sec:distances}, 
the most fundamental one, that requires the least structure is the
\emph{geometric distance} $\lambda$; since it only uses the information of the
4-velocity of the observer and the null geodesic coming from the source.
In fact, it
can be applied to any spacetime and any past directed null geodesic
and is the dynamical parameter of the fundamental equation (\ref{eq. dev geo}).
Consequently, it constitutes the
natural notion of distance in such context. 

Then, for a source of size $\mathcal{X}$ at a distance $\lambda$ along the past null 
cone this observer would expect to see, in the absence of gravitational effects, 
that the source is subtending an angle $\delta 
\beta$:
\begin{equation}
\delta \beta = \frac{\mathcal{X}}{\lambda}.
\end{equation}
However, in R-W geometries equation (\ref{eq: X solution}) tells us that 
\begin{equation}\label{eq:chideteta}
\mathcal{X} = A(t) f_k(\chi) \delta \theta .
\end{equation}
Then, from equation (\ref{eq:D-A-fisrt-defin}) one has the simple relation
\begin{equation}\label{eq:dbetaDAlambdadteta}
\delta \beta = \frac{D_A}{\lambda} \delta \theta .
\end{equation}

The relation between the observed angle $\delta \theta$ and $\delta \beta$ 
is understood as the lensing effect produced by the R-W cosmology. 
In the usual language of weak 
lensing this constitute a very peculiar lens since the whole spacetime 
acts as a lens which is not placed 
at a particular distance. 
Furthermore, in the R-W gravitational lens there is no bending angle due 
to the fact that it is an homogeneous and isotropic spacetime. In spite of this, one 
still has a non-trivial meaning of the optical scalars.

It is important to remark this fact because it is a source of confusion.
For instance, the diagram appearing in figure \ref{f:fig0} comes from the normal 
situation one encounters for a localized gravitational lens. However, in the case in which
the whole spacetime is acting as a lens (which is not localized) the diagram must be
understood in the conceptual way using the definitions of $\delta \beta$ and $\delta \theta$ 
as the angles measured with no lens, and with lens, respectively.
This issue is completely missed in approaches based on the bending angle
concept, as is normally done in textbooks\cite{Schneider92,Schneider06}.

Another essential point that one should also note is that
equations (\ref{eq:chideteta}) and (\ref{eq:dbetaDAlambdadteta})
are exact equations valid also for large values of $\delta \theta$.
This means that one can regard the R-W lens in a non-perturbative
way \cite{Frittelli00a, Frittelli00}.

\subsection{The cosmic convergence and angular cosmic magnification}

Equation (\ref{eq:dbetaDAlambdadteta}) is telling us that no cosmic shear is
present, and that the \emph{cosmological convergence} $\kappa_c$ 
can be deduced from
\begin{equation}\label{eq:betacosmic}
\delta \beta = \left( 1 - \kappa_c \right)  \delta \theta ;
\end{equation}
so that one has
\begin{equation}\label{eq:kappadedaylambda}
\kappa_c  = 1 - \frac{D_A(\lambda)}{\lambda}.
\end{equation}
We emphasize again that the cosmic convergence $\kappa_c$ can not be
defined using the standard approach to gravitational lenses
bases on the bending angle concept.

In general, $D_A(\lambda) < \lambda$ holds for all values of $\lambda$ and for 
close objects one has that $\kappa_c \ll 1$. In fact, this is the common situation 
that one finds in the discussion of weak gravitational lenses in which the magnitudes of 
the optical scalars are always much smaller than the unit value and eq. 
(\ref{eq:standarlensequation}) represents an approximation for small variations from 
chosen angular directions. 
However as indicated above, in the R-W geometry the cosmic convergence
is exactly given by (\ref{eq:kappadedaylambda}), which goes to the unit value
as one approaches the initial cosmic singularity.

The behaviour of $\kappa_c$ close to the observer can be investigated taking an 
expansion of $D_A$ in terms of $\lambda$; we found:
\begin{equation}
\kappa_c = \kappa_c^{(2)}\lambda^2 + \kappa_c^{(3)}\lambda^3 + \mathscr{O}
\left(\lambda^4\right);
\end{equation}
where
\begin{align}
\kappa_c^{(2)} &= \left. \frac{1}{3!}\Phi_{00} \right|_{\lambda = 0} ,\\
\kappa_c^{(3)} &= \left. -\frac{2}{4!} \ell \left( \Phi_{00} \right) 
\right|_{\lambda = 0}.
\end{align}

% se saco de aqui **************************

From the previous discussion of angular magnification in general spacetimes, 
one can now see that the \emph{cosmic magnification} $\mu_c$, is given by
\begin{equation}\label{eq:mu-cosm}
\mu_{c} = \frac{1}{\left(1 - \kappa_{c}\right)^2}
= \left(\frac{\lambda}{D_A(\lambda)}\right)^2
.
\end{equation}

It is probably worthwhile to emphasize that the content of the previous
equation coincides with equation (\ref{eq:dbetaDAlambdadteta}); which
as mentioned before it is an exact equation valid even for large
values of $\delta \theta$.

When the previous theorem, on the relation between magnification and intensity
magnification, is applied to the cosmological scenario, one obtains: 
\begin{equation}
\begin{split}
\tilde{\mu}_c 
%= \frac{1}{\left(1 - \kappa_{c}\right)^2}
%= \left(\frac{\lambda }{D_A(\lambda)} \right)^2
= \mu_c 
.
\end{split}
\end{equation}

Let us notice that $\mu_c$ diverges as the source is considered close to the 
initial singularity; which in turns indicates that $\kappa_c$ approaches the 
unit value, and therefore it can not be assumed to be small at large distances.

\subsection{Example 1: Milne Universe}
As a simple illustration of the family of cosmological ~R-W lenses we 
consider the simplest one which corresponds to the Milne Universe;
which is the case of an empty cosmology,  
which is interpreted as the limit of vanishing density of an open dust model.
For these reason, the whole Milne Universe, defined by the coordinates of the R-W metric, 
is identical from the point of view of the geometry to a portion of Minkowski spacetime.
This geometry is characterized by $k=-1$ and $A(t) = ct$, since we are assuming $A$
has units of length; although we use units for which the universal constant
$c$ is one. 
In terms of co-moving coordinates
the relevant distances for the discussion of lenses are
\begin{equation}
\lambda_\text{Milne} = \frac{1}{2}\frac{A^2(t_0)- A^2(t)}{A(t_0)} 
%= \frac{1}{2} \frac{t_0^2- t^2}{t_0} 
= A(t_0)\,  z \, \frac{1 + \frac{z}{2}}{(1+z)^2}
, 
\end{equation}
\begin{equation}
D_A = \lambda , 
\end{equation}
\begin{equation}
D_L = \left( \frac{A(t_0)}{A(t)}\right)^2 \lambda ; 
\end{equation}
which then yield a vanishing cosmological convergence as expected:
\begin{equation}
\kappa_c = 0, 
\end{equation}
\begin{equation}
\mu_c = \tilde{\mu}_c = \mu'_c = 1. 
\end{equation}

\subsection{Example 2: Friedman Universes}\label{subsec:Fried-Universe}

We concentrate in this subsection on the non-flat(4-dimensional sense) 
Friedman Universes;
in particular we will focus on the models that involve only dust, 
radiation and cosmological constant.

In general, analytical expressions for the optical quantities can not be obtained,
however we content our self with the study of the behaviour in a neighbourhood of a 
fundamental observer.

The curvature component $\Phi_{00}$ and its derivative with respect to $\ell$ have the
expressions:
\begin{align}
\Phi_{00} &= \frac{4\pi G}{c^2} \frac{A^2(t_0)}{A^2(t)}  \left( \varrho(t) + 
\frac{P(t)}{c^2} \right); \\
\begin{split}
\ell\left(\Phi_{00}\right) &= \frac{4\pi G}{c^3}\frac{A^3(t_0)}{A^3(t)} 
\bigg[ \frac{d\varrho}{dt} 
+ \frac{1}{c^2}\frac{dP}{dt} \\
&\;\;\;\; - 
2H(t) \bigg( \varrho(t) + \frac{P(t)}{c^2}  \bigg)  \bigg];
\end{split}
\end{align}
where $G$ is the gravitational constant, $c$ is the speed of light,
$\varrho(t)$ and $P(t)$ the matter density and
pressure of the cosmic fluid respectively and $H(t)$ is the Hubble's rate of expansion 
defined as
\begin{equation}
H(t) \equiv \frac{1}{A(t)}\frac{dA}{dt}.
\end{equation}
For convenience in this section and in all our final results we
show the appearance of the Universal constants.

We can write all quantities of the fluids and the geometric functions,
as for example $H(t)$, in terms of the so-called 
\emph{density parameters} $\left( \Omega_m, \Omega_r, 
\Omega_{\Lambda} \right)$
and the \emph{critical density} $\varrho_{cr}$:
\begin{align}
\varrho(t) &= \varrho_{cr}\bigg( \Omega_m \frac{A^3(t_0)}{A^3(t)} + 
\Omega_r \frac{A^4(t_0)}{A^4(t)} + \Omega_{\Lambda} \bigg), \\
\frac{P(t)}{c^2} &= \varrho_{cr}\bigg( \frac{\Omega_r}{3} \frac{A^4(t_0)}{A^4(t)} - 
\Omega_{\Lambda} \bigg), 
\end{align}
%\verde{
\begin{equation}
\label{eq:friedman}
\frac{H^2(t)}{H^2(t_0)} =
-\frac{k c^2}{H(t_0)^2 A(t)^2} %  \\
+ \Omega_m \frac{A^3(t_0)}{A^3(t)} + \Omega_r \frac{A^4(t_0)}{A^4(t)} +\Omega_{\Lambda} \; .
\end{equation}
The evaluation of $A(t_0)$ is through the equation
\begin{equation}
1 = 
-\frac{k c^2}{H(t_0)^2 A(t_0)^2}
+ \Omega_r  + \Omega_m + \Omega_{\Lambda}  ;
\end{equation}
which is just the evaluating of (\ref{eq:friedman}) at the present time.
In the case $k=0$ one has a freedom in the value of $A(t_0)$ which can be taken 
as $A(t_0) = \frac{c}{H(t_0)}$.
%}

This yields,
\begin{align}
\Phi_{00} &= \frac{4\pi G \varrho_{cr}}{c^2} \frac{A^5(t_0)}{A^5(t)} 
\left(\Omega_m  + \frac{4}{3}\Omega_r \frac{A(t_0)}{A(t)} \right),
\\
\ell\left(\Phi_{00}\right) &= - \frac{4\pi G \varrho_{cr}}{c^3}\frac{A^6(t_0)}{A^6(t)} H(t) 
\left(5 \Omega_m + 8 \Omega_r \frac{A(t_0)}{A(t)} \right).
\end{align}

Then, the cosmological convergence $\kappa_c$, near the observer, 
behaves in the following way:
\begin{equation}
\begin{split}
\kappa_c \left( \lambda \right) &= \frac{4\pi G \varrho_{cr}}{3! c^2}
\bigg( \Omega_m + \frac{4}{3}\Omega_r \bigg)\lambda^2 \\
& + \frac{8\pi G \varrho_{cr}}{4! c^3} H_0 \bigg( 5\Omega_m + 8 \Omega_r \bigg)\lambda^3
+ \mathscr{O}\left( \lambda^4 \right).
\end{split}
\end{equation}

It is probably interesting to remark that the only optical scalar that appears
in the gravitational lens study of the R-W geometry is independent
of the cosmological constant contribution $\Omega_\Lambda$.

\subsection{Typical values for the cosmic convergence and magnification functions}

The last examples allow  to estimate the effect of the R-W spacetimes
as a gravitational lens in a simple way.
In this section we 
present the values of the cosmic convergence and magnification 
functions  for a couple of representative cosmic models; one based on 
the data of the Planck Collaboration and the other just coming from
the primordial nucleosynthesis calculations.

\subsubsection{Values from Planck Collaboration}

The following values, used and reported by the Planck Collaboration
in reference \cite{Ade:2013zuv},
were employed to compute the different quantities 
associated to the lens effect; that we show below in table \ref{tab:Planck-2}.
\begin{align}
\Omega_m &= 0,314 \pm 0,020, \\
\Omega_{\Lambda} &= 0,686 \pm 0,020 , \\
H_0 &= 67,4 \pm 1,4 \, \frac{\text{km}}{\text{s} \cdot \text{Mpc}}, \\
k &= 0 . 
\end{align}

\begin{table}%[H]
	\caption{Planck $\Lambda$CDM model}\label{tab:Planck-2}
%	\begin{ruledtabular}
		\begin{tabular}{cccccc}
			$\lambda$ $\left[ \text{Mpc} \right]$ & $z_0$ & $z$ & $\kappa_c$ & $\tilde{\mu}_c=\mu_c$  & $\mu'_c$  \\
			\hline
			43.95   & 0.010 & 0.01 & 7.85 $\times$ $10^{-6}$    & 1.00002 & 0.99481 \\
			169.55  & 0.040 & 0.04 & 0.00013                    & 1.00025 & 0.98043 \\
			395.164 & 0.103 & 0.10 & 0.00078                    & 1.00157 & 0.95642 \\
			591.27  & 0.167 & 0.16 & 0.00200                    & 1.00401 & 0.93785 \\
			788.85  & 0.245 & 0.23 & 0.00411                    & 1.00826 & 0.92189 \\
			935.68  & 0.314 & 0.29 & 0.00647                    & 1.01307 & 0.91231 \\
			1179.06 & 0.459 & 0.41 & 0.01266                    & 1.02581 & 0.90235 \\
			1311.31 & 0.561 & 0.49 & 0.01776                    & 1.03649 & 0.90127 \\
			1474.92 & 0.722 & 0.61 & 0.02668                    & 1.05557 & 0.90624 \\
			1735.04 & 1.131 & 0.88 & 0.05107                    & 1.11055 & 0.93905 \\
			1915.09 & 1.680 & 1.18 & 0.08261                    & 1.18821 & 0.99991 \\
			2121.16 & 3.629 & 1.85 & 0.15795                    & 1.41034 & 1.19321  
		\end{tabular}
%	\end{ruledtabular}
\end{table}

In the above list we have also included the corresponding values for $z_0$; 
i.e. the value that would take the redshift if the universe where 
the flat Milne spacetime, for a given geometric distance $\lambda$.

Then, it is curious that while the convenient astrophysical intensity magnification $\mu'_c$
shows a fainter behavior for this range of redshift values,
in qualitative agreement with
observations\cite{Riess:1998cb,Perlmutter:1998np,Riess:2004nr},
the more physical intensity magnification $\tilde\mu_c$ implies
a brighter behavior for the same range.
This emphasizes the fact that the notions 
of fainter or brighter, which are very common in the observational language, are rather 
relative; since they depend on what setting one is assigning these concepts.

\subsubsection{Values from primordial nucleosynthesis calculations}
Here we consider the set of parameters that come just from the primordial nucleosynthesis 
calculations and independent observation of the Hubble parameter, and others,
that are shown next. 
\begin{align}
\Omega_m &= 0,042569  , \\
\Omega_r &= 4.7647 \times 10^{-5} , \\
\Omega_{\Lambda} &= 0  , \\
H_0 &= 72 \, \frac{\text{km}}{\text{s} \, \text{Mpc}}, \\
k &= -1 .
\end{align}

In table \ref{tab:Baryonic} below, we present 
the result of the calculations for the different quantities associated 
to the cosmic lens effects, using these values.
\begin{table}%[H]
	\caption{Baryonic Low density model }\label{tab:Baryonic}
%	\begin{ruledtabular}
		\begin{tabular}{ccccccc}
			$\lambda$ $\left[ \text{Mpc} \right]$ & $z_0$ & $z$ & $\kappa_c$ & $\tilde{\mu}_c=\mu_c$ & $\mu'_c$ \\
			\hline
			41.04   & 0.010 & 0.01 & 1.06 $\times$ $10^{-6}$ & 1.00000 & 1.00021 \\
			157.07  & 0.040 & 0.04 & 0.00002 				 & 1.00003 & 1.00087 \\
			361.12  & 0.100 & 0.10 & 0.00010                 & 1.00020 & 1.00223 \\
			560.36  & 0.170 & 0.17 & 0.00028                 & 1.00057 & 1.00391 \\
			748.05  & 0.249 & 0.25 & 0.00059                 & 1.00118 & 1.00592 \\
			884.86  & 0.319 & 0.32 & 0.00093                 & 1.00187 & 1.00776 \\
			1114.46 & 0.467 & 0.47 & 0.00189                 & 1.00378 & 1.01191 \\
			1232.04 & 0.565 & 0.57 & 0.00266                 & 1.00534 & 1.01483 \\
			1379.10 & 0.720 & 0.73 & 0.00409                 & 1.00824 & 1.01969 \\
			1619.77 & 1.121 & 1.15 & 0.00872                 & 1.01767 & 1.03341 \\
			1793.70 & 1.684 & 1.77 & 0.01701                 & 1.03491 & 1.05540 \\
			1913.14 & 2.503 & 2.77 & 0.03215                 & 1.06754 & 1.09353
		\end{tabular}
%	\end{ruledtabular}
\end{table}

As expected, this model which is practically devoid of matter,
presents a behaviour with small deviations from the trivial flat Milne 
Universe.
We also note that the intensity magnification $\mu'_c$ is always 
greater than one in the whole range of the affine parameter.

\subsubsection{Plots for these previous models}

In this subsection we present a graph illustrating the behaviour of 
the physical intensity magnification $\tilde \mu_c = \mu_c$
in terms of the geometric distance,
together with the graph of the intensity magnification $\mu'_c (z)$;
as used in astrophysical works.
\begin{figure}%[H]
	\centering
	\includegraphics[clip,width=0.45\textwidth]{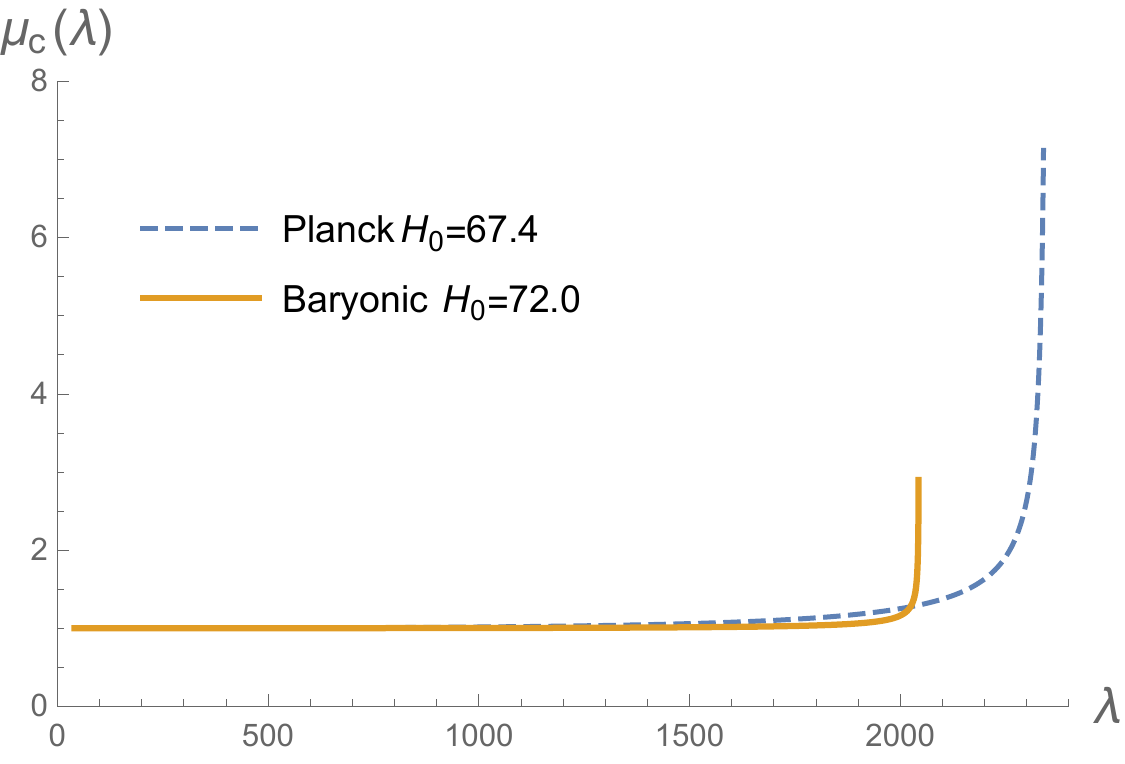}
	\caption{
		Since in general $\lambda > D_A(\lambda)$ the cosmic
		magnification is greater than one and divergent near of the initial 
		singularity, except in the trivial case of flat spacetime. 
	}
	\label{fig:Mu-Cosm-2}
\end{figure}
\begin{figure}%[H]
	\centering
	\includegraphics[clip,width=0.45\textwidth]{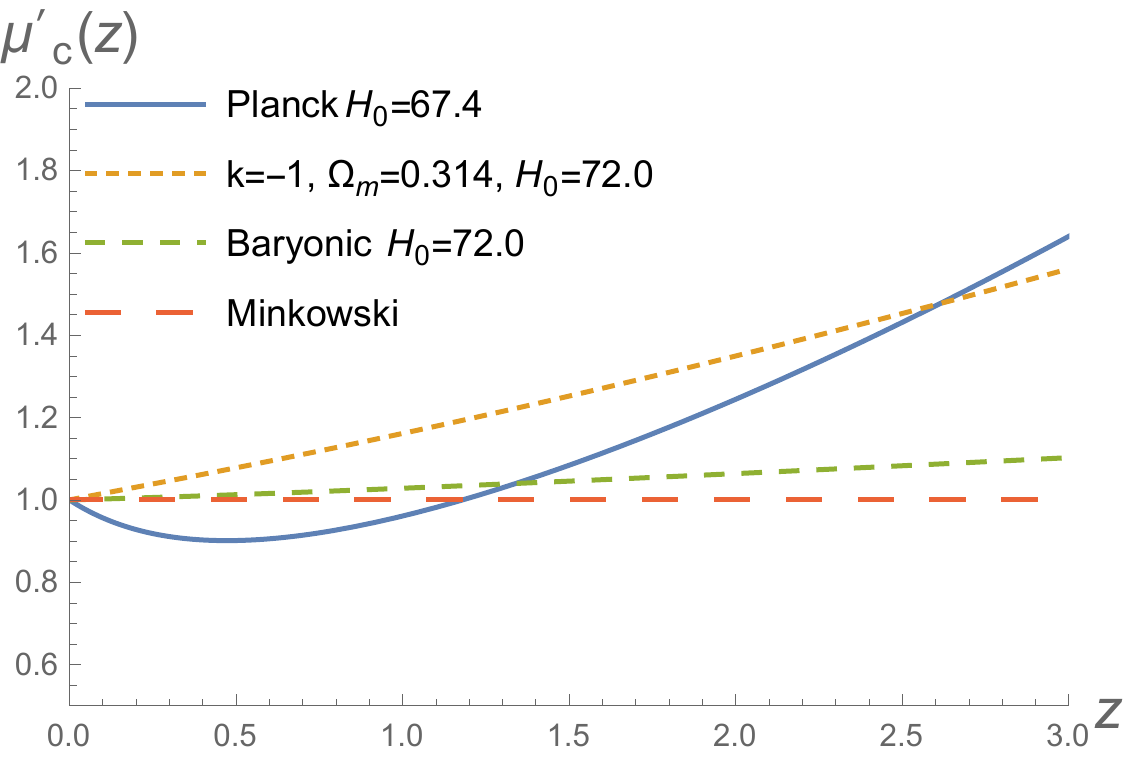}
	\caption{
		The astrophysical magnification $\mu'_c(z)$ 
		is shown for the Planck and Baryonic models.
		Milne Universe and an open universe with only matter content 
		$\Omega_m = 0.314$ are also shown.
	}
	\label{fig:Mu-CosmPrima}
\end{figure}
One can see in these graphs how the notion of `brighter' and/or `fainter'
depends in the way the observational data is studied.

\section{The presence of an additional lens over the cosmology}\label{sec:Addit-lens}

\subsection{The main exact equation}

In the previous section we have shown that a R-W spacetime can be thought
as a weak gravitational lens. In the current section we will deal with
additional lenses on the geometry. 
We mean by this an alteration of the completely
homogeneous and isotropic cosmological background where the size of the 
additional lens must be much smaller that the cosmological scale, since otherwise 
it would change the cosmology one is discussing.
This means that normally the time of flight of photons trough the additional lens 
would be considered small compared with the total time of flight.
By time of flight we mean more precisely the lapse of the geometric affine
distance.

In order to compute distortions due to the presence of inhomogeneities we will
consider the exact equation (\ref{eq. dev geo}) 
with the source of curvature component
decomposed in terms of the cosmological background 
and the lens contribution,
in the following way:
\begin{equation}\label{eq: dev geo pert0}
\ell\left(\ell \left(\boldsymbol{\mathcal{X}} \right) \right) = 
- \left( Q_B + Q_L  \right)\boldsymbol{\mathcal{X}};
\end{equation}
where we have included the subindex $B$ to denote intrinsic quantities of the 
background and 
$L$ to denote quantities associated to the additional lens.
In other words, the components of the curvature appearing in $Q$
in equation (\ref{eq. dev geo})	is now expressed as
\begin{equation}\label{eq:Q-decomposition}
Q = Q_B + Q_L .
\end{equation}
In particular $Q_L$ is the matrix containing the difference from 
the background curvature which is given in terms of the 
scalars:
\begin{equation}
Q_L = \begin{pmatrix}
\Phi^L_{00} & \Psi^L_0 \\
\bar{\Psi}^L_0   & \Phi^L_{00}
\end{pmatrix}.
\end{equation}

%\verde{
As mentioned in the introduction, it is worthwhile to
emphasized that in the decomposition of the total curvature
in terms of the background plus a lens term, it is not assumed
that the additional term must be small in any sense.
That is we are not assuming at this
stage any kind of perturbation; it is just the representation
of an exact geometry, without perturbation, 
in terms of a decomposition with respect to a chosen background.
This becomes more clear when we fix the issue of gauges
in the next subsection.
%}

\subsection{A treatment of the main equation without room for gauges}

As usual, referring quantities with respect to a background, lead us to 
deal with the delicate issues associated to the gauge freedom.
So, with the aim to clarify the way in which equation 
(\ref{eq:Q-decomposition}) should be understood, 
it is pertinent to present a short explanation
before tackling the problem of solving the equation.

A gauge choice in the sense of \cite{Stewart:1974uz, Stewart:1990fm, Bardeen:1980kt}
is a correspondence between a fiducial background (namely a R-W one) and
the real inhomogeneous spacetime.
In the context of cosmological perturbations the problem of how to specify this map 
appropriately has been called the ``fitting problem''\cite{Ellis:1987zz}.
A good understanding of the way in which one fully specifies the gauge is crucial at 
the moment to interpret tensor fields quantities that are regarded as departure
from the background\cite{Ellis:1989jt}.

In the simpler case of weak lensing over a flat background, 
the meaning of $Q_L$ is always well defined 
since the curvature is  gauge invariant in such case.
It is because of this specific fact that in reference\cite{Gallo11} one prefers 
to express the optical scalar in terms of the curvature. 

The discussion of gauge invariance is subtle, and one should have
a clear picture of the geometrical and physical framework one is using.
For example in references \cite{Stewart:1974uz, Stewart:1990fm} they discuss
the notion of \emph{identification gauge invariant}(i.g.i)	quantities.
Then, since i.g.i. quantities should vanish on the background, one would deduce
that while $\Psi^L_{0}$ is i.g.i, the curvature scalar $\Phi^L_{00}$ is not.
However their treatment does not apply to our setting, as we now explain.

In our work, all these gauge issues are solved when the curvature
components $Q_B$ is taken from observation.
The measurements we have in mind are all the cosmological observations,
as for example Plack's CMB measurement, or the supernovae
observations, mentioned in references \cite{Perlmutter:1998np,Riess:1998cb},
where researchers used the cosmological observation of
luminosity and redshift of supernovae to fit a Robertson-Walker
geometry to the actual spacetime.
After the parameters determining the Robertson-Walker geometry are
fixed from observation, and the direction and magnitude of our
velocity with respect the CMB is determined; the RW geometry
is completely determined in our past null cone;
this means metric, connection and curvature.	
In other words, the RW cosmic background is uniquely defined
on our past null cone in terms of observations;
and therefore there is no gauge freedom allowed in this respect.

Then, since (\ref{eq: dev geo pert0}) is an exact equation in terms
of the exact geodesics, and $Q$ is therefore unambiguously defined;
one concludes that $Q_L$ also does not have room for any gauge freedoms.
In our framework $Q_L$ is the contribution to the curvature that is needed
to explain the gravitational lensing observations.
That is, both curvature terms $Q_B$ and $Q_L$ are determined from
independent set of observations; where no gauge freedoms
intervene.

Since one has at hand the back ground RW geometry, one can also
decompose the exact metric in terms of $g^{ab} = g^{ab}_{B} + h^{ab}$;
where $g^{ab}_{B}$ is the metric of the RW geometry, and 
$h^{ab}$ is the needed tensor to complete the metric.
Therefore we have at our disposal the complete RW geometry,
including metric and solutions of geodesic equations;
that we will use next.
We should probably remark that at this stage we do not assume
a particular form for $h^{ab}$, since we only require
to describe the observed $Q_L$.

\subsection{The technique for solving the main equation}

Let us mention again that equation (\ref{eq: dev geo pert0}) is exact,
and therefore it uses the exact geodesics; however, in order to find solutions 
it is more convenient to carryout the calculation, at the linear order in the 
curvature $Q_L$, using the background geodesics.
This forces us to estimate the magnitude of the error involved in this procedure.
So, in the following we present an estimate of this error.

Let us begin denoting by $\boldsymbol{\mathcal{X}}_E$ the exact solution along the exact 
null geodesic, which we indicate as $x_E^\alpha(\lambda)$. 
Then, the integral version of equation (\ref{eq: dev geo pert0}) is
\begin{equation}\label{eq:X-Exact-Solution}
\begin{split}
\boldsymbol{\mathcal{X}}_E& (\lambda) = \ell(\boldsymbol{\mathcal{X}}_E) \parallel_{0} 
\left( \lambda - \lambda_0 \right) \\
& - 
\int_{\lambda_0}^{\lambda} \int_{\lambda_0}^{\lambda'}\left( Q_B + Q_L \right)(x_E^{\alpha}(\lambda'')) 
\boldsymbol{\mathcal{X}}_E(x_E^{\alpha}(\lambda'')) d\lambda'' d\lambda' .
\end{split}
\end{equation}
Now, let us also consider the above expression computed along the geodesics of the 
background which we will indicate by $x^{\alpha}(\lambda)$; we mean
\begin{equation}\label{eq:X-along-background}
\begin{split}
\boldsymbol{\mathcal{X}} (\lambda) =& \ell(\boldsymbol{\mathcal{X}}) \parallel_{0} 
\left( \lambda - \lambda_0 \right) \\
& - 
\int_{\lambda_0}^{\lambda} \int_{\lambda_0}^{\lambda'}\left( Q_B + Q_L \right)(x^{\alpha}(\lambda'')) 
\boldsymbol{\mathcal{X}}(x^{\alpha}(\lambda'')) d\lambda'' d\lambda' ;
\end{split}
\end{equation}
where in order to differentiate the result from the exact one, we have omitted the subindex $E$ for
the vector $\boldsymbol{\mathcal{X}}(x^{\alpha}(\lambda))$ which contains the components of the 
deviation vector (\ref{eq. dev vector}).
To simplify the notation we are using the same symbols for the curvature components
but it should be remarked that there is a slight change in the null tetrads.
The relation between the exact geodesics and those of the background will be denoted 
as $x^{\alpha}_E(\lambda) - x^{\alpha}(\lambda) \equiv \delta x^{\alpha}(\lambda)$.

Then, we will consider the difference 
$\boldsymbol{\delta} \boldsymbol{\mathcal{X}}(x^{\alpha}(\lambda)) 
\equiv 
\boldsymbol{\mathcal{X}}_E(x^{\alpha}_E(\lambda)) -
\boldsymbol{\mathcal{X}}(x^{\alpha}(\lambda))$ between the exact result, namely equation (\ref{eq:X-Exact-Solution}) 
and the alternative procedure given by equation (\ref{eq:X-along-background}) and we will show that
such difference is lower order than $\boldsymbol{\mathcal{X}}(\lambda)$. 
Defining $\boldsymbol{\delta} \boldsymbol{\mathcal{X}}$ in this way it is 
implicit that we are using the same procedure to define the auxiliary vector
$m$ and $\bar m$ for both geodesics.
Notice that $\delta$ quantities are of order $\mathscr{O}\left( Q_L \right)$.
Explicitly, we have
\begin{equation}
\begin{split}
\boldsymbol{\delta} \boldsymbol{\mathcal{X}} =& - 
\int_{\lambda_0}^{\lambda} \int_{\lambda_0}^{\lambda'}\left( Q_B + Q_L \right)(x_E^{\alpha}(\lambda'')) 
\boldsymbol{\mathcal{X}}_E(x_E^{\alpha}(\lambda'')) d\lambda'' d\lambda' \\
&+ 
\int_{\lambda_0}^{\lambda} \int_{\lambda_0}^{\lambda'}\left( Q_B + Q_L \right)
(x^{\alpha}(\lambda'')) 
\boldsymbol{\mathcal{X}}(x^{\alpha}(\lambda'') ) d\lambda'' d\lambda' \\
=& - 
\int_{\lambda_0}^{\lambda} \int_{\lambda_0}^{\lambda'}\Big[ Q_B (x_E^{\alpha}(\lambda'')) 
- Q_B (x^{\alpha}(\lambda'')) \Big]
\boldsymbol{\mathcal{X}}(x^{\alpha}(\lambda'')) d\lambda'' d\lambda' \\
& - 
\int_{\lambda_0}^{\lambda} \int_{\lambda_0}^{\lambda'}\Big[ Q_L (x_E^{\alpha}(\lambda'')) 
- Q_L (x^{\alpha}(\lambda'')) \Big]
\boldsymbol{\mathcal{X}}(x^{\alpha}(\lambda'')) d\lambda'' d\lambda' \\
&- 
\int_{\lambda_0}^{\lambda} \int_{\lambda_0}^{\lambda'}\left( Q_B + Q_L \right)
(x_E^{\alpha}(\lambda'')) 
\boldsymbol{\delta} \boldsymbol{\mathcal{X}}(x^{\alpha}(\lambda'') ) d\lambda'' d\lambda' ;
\end{split}
\end{equation}	
where the initial derivatives of $\boldsymbol{\mathcal{X}}$ do not appear since, they represent 
the values of the observations, that are the same for both calculations; in other words,
we are here comparing two calculation for the optical scalars in terms of
curvature $Q_L$ that is determined by the existence of the lens.
Also, let us remember that in weak lensing computations one is interested in the first order 
effects of the curvature $Q_L$.
We will see below that the expressions for $\boldsymbol{\mathcal{X}}$ will involve
terms up to order $\mathscr{O}\left( Q_B Q_L \right)$.
In any case, one can see that just keeping linear terms in $Q_L$, from
the previous expression one obtains:
\begin{equation}
\begin{split}
\boldsymbol{\delta} \boldsymbol{\mathcal{X}} =& - 
\int_{\lambda_0}^{\lambda} \int_{\lambda_0}^{\lambda'} \\
&\Big[ Q_B (x_E^{\alpha}(\lambda'')) 
- Q_B (x^{\alpha}(\lambda'')) \Big]
\boldsymbol{\mathcal{X}}(x^{\alpha}(\lambda''))  d\lambda'' d\lambda' \\
&- 
\int_{\lambda_0}^{\lambda} \int_{\lambda_0}^{\lambda'} 
Q_B(x^{\alpha}(\lambda'')) 
\boldsymbol{\delta} \boldsymbol{\mathcal{X}}(x^{\alpha}(\lambda'') ) d\lambda'' d\lambda' \\
& + \mathscr{O}\left( Q_L^2 \right);
\end{split}
\end{equation}	
where only $Q_B$ appears explicitly.

First of all, let us note that $\boldsymbol{\delta} \boldsymbol{\mathcal{X}}$
appears only in the second term of the right hand side; so that 
a first estimate of it is just the first term.
Then, let us also see that the exact null vector $\ell$ can always be expressed with respect
to an appropriately chosen null geodesic vector of the background
with a slight correcting conformal factor.
To this, let us add that although $Q_B$ refers to an homogeneous spacetime,
both geodesics are different in the lens and therefore have different
affine length in the lens.
Both effects are of the same nature and can be estimated by the difference
in the affine length of the geodesics inside the lens.
Let us call this difference
$\delta \lambda_L$.
If $\alpha$ is the deviation angle for a typical geodesics; then
one could estimate $\delta \lambda_L$
from $\delta \lambda_L = \delta _l (1- \cos(\alpha))$ 
where $\delta_l$ is the characteristic radial size of the lens.
Let us apply this to a spherical distribution of matter,
and to a photon trajectory with impact parameter $J$,
such that a quantity $M$ of matter is contained in a
sphere of radius $J$; then one would have $\alpha = \frac{4 M}{J}$,
using geometric units.
For a typical $\alpha \ll 1$ one would have
$\delta \lambda_L \cong \delta _l \frac{\alpha^2}{2} = \delta _l \frac{8 M^2}{J^2}$.
Proceeding with the calculation we now estimate
\begin{equation}
\begin{split}
- 
\int_{\lambda_0}^{\lambda} \int_{\lambda_0}^{\lambda'} &
\Big[ Q_B (x_E^{\alpha}(\lambda'')) 
- Q_B (x^{\alpha}(\lambda'')) \Big]
\boldsymbol{\mathcal{X}}(x^{\alpha}(\lambda''))  d\lambda'' d\lambda' \\
&\cong
(\lambda_s - \lambda_l ) Q_B(\lambda_l) \boldsymbol{\mathcal{X}}(x^{\alpha}(\lambda_l) )
\delta \lambda_L \\
&=
(\lambda_s - \lambda_l ) Q_B(\lambda_l) \boldsymbol{\mathcal{X}}(x^{\alpha}(\lambda_l) )
\delta _l \frac{8 M^2}{J^2}
.
\end{split}
\end{equation}	
Typically we will take $J \approx \frac{\delta_l}{2}$.
Let us apply this to a typical situation in which the lens
has a radshift $z_l = 0.06$, while the lens has redshift of $z_s=0.33$,
for a galaxy we take $\delta_l \sim 200 \text{kpc}$ and $M \sim 10^{-8}\text{Mpc}$;
while for a cluster of galaxies we take
$\delta_l \sim 1 \text{Mpc}$ and $M \sim 10^{-5}\text{Mpc}$.
Then we obtain for a galaxy 
$\frac{\boldsymbol{\mathcal{\delta X}}}{\boldsymbol{ \mathcal{X}}} \sim 10^{-11}$,
and for a cluster of galaxies
$\frac{\boldsymbol{\mathcal{\delta X}}}{\boldsymbol{ \mathcal{X}}} \sim 10^{-6}$.

We conclude then that using the null geodesics of the cosmological background
to carry out the calculation involves deviations of the values that are
completely negligible; and therefore it is safe to use this technique.

After corroborating the feasibility of the path integration technique, let us concentrate
in the details of how to handle the algebra of the computation.
Let us recall that
from the physical point of view, one integrates equation (\ref{eq: dev geo pert0})
from the position of the observer along the past null cone.
Therefore, although one could start with finite values of $\mathcal{X}$,
we are only interested on initial conditions that have the information
of the null cone, and so, our $\mathcal{X}$'s start with a diverging behavior;
dictated by the observing angle.
For this reason we look for solutions of the form
\begin{equation}
\boldsymbol{\mathcal{X}} = M D_A \boldsymbol{\mathcal{V}}_0
\end{equation}
where $M$ is a well behaved $2 \times 2$ square matrix and $\boldsymbol{\mathcal{V}}_0$ has
the same meaning than before; it is the observed angle subtended by the image:
$\boldsymbol{\mathcal{V}}_0 \equiv \delta \theta$.
This ansatz is reminiscent of the 
solution (\ref{eq: X solution}), since one recognizes the 
background solution
$\boldsymbol{\mathcal{X}}_B = D_A \boldsymbol{\mathcal{V}}_0$.

It is probably interesting to point out that in the weak lensing regime, the matrix
$M$ can be related to the optical scalars matrix $\mathcal{A}$
mentioned in (\ref{eq:standarlensequation}); since  taking
the relation $\lambda \,\delta\beta = \mathcal{X}$ one has
\begin{equation}\label{eq:Matrix-A}
\mathcal{A} 
= \frac{D_A}{\lambda} \, M
= \left(1 - \kappa_c \right) \, M .
\end{equation}
In other words, the total gravitational lens combines the global cosmic convergence 
$\kappa_c$ with the `local' lensing effects contained in $M$.

In writing equation (\ref{eq:Matrix-A}) we should note that we have change
the basis in which the matrix $\mathcal{A}$ is represented; since
the matrix $\mathcal{A}^a_{\; b}$ of equations (\ref{eq:standarlensequation})
and (\ref{eq:optical-scalars0}) refer to a base in the sphere of
observed directions, while $\mathcal{A}$ in (\ref{eq:Matrix-A}) 
refers to the complex components of $\boldsymbol{\mathcal{X}}$.
In order to relate both representations it is necessary to give
an explicit relation of the complex vectors $m$ and $\bar m$ in terms
of directions that can be expressed in turn in terms of 
directions in the sphere of observation.
To fix ideas, let us consider the expression (\ref{eq:dL-standard3}) 
of the space line element and consider the beam of light is coming from
the positive $y$ direction, of a conformal Cartesian frame.
Then, at any point we have at our disposal an orthonormal frame
which respects the conformal Cartesian frame directions, and we
express the null tetrad adapted to the incoming null geodesic by:
\begin{equation}\label{eq:tetradflat}
\begin{split}
l^a=&(1,0,-1,0),\\
m^a=&\frac{1}{\sqrt{2}}(0,1,0,i),\\
\bar{m}^a=& \frac{1}{\sqrt{2}}(0,1,0,-i),\\
n^a=&\frac{1}{2}(1,0,1,0) ;
\end{split}
\end{equation}
which is a slight different notation that the one used in \cite{Gallo11}.

Now, in order to compare with the usual expressions for the lens scalars 
$\kappa, \gamma_1$ and $\gamma_2$, 
let us recall that they are defined via the relations 
(\ref{eq:standarlensequation}) and (\ref{eq:optical-scalars0});
but since it is a linear relation, one can relate the deviation vectors by the same
matrix, namely
\begin{equation}\label{eq:deltabeta}
\varsigma^i_s=A^i_j\varsigma^j_o;
\end{equation}
where $\{\varsigma^i_s,\,\varsigma^i_o\}$ are the spatial vector associated with
$\{\varsigma_s,\,\varsigma_o\}$ respectively.
In this expression,
it is needed to determine the meaning of the indices $(i,j)$ of the two dimensional space
of the images. 
For the purpose of connecting
the natural Cartesian orientation we identify the first
component of the two dimensional space with the $x$ direction, 
and the second component with the $z$ one.
We need then, to know the components of the spatial vectors $\varsigma^a_o$ generated 
by $\varsigma_o$ and similarly by $\varsigma_s$ in a Cartesian like coordinate system.
In the case of $\varsigma^a_o$, it is given by
\begin{equation}
\begin{split}
\varsigma^a_o=&\varsigma_o \bar{m}^a+\bar\varsigma_o{m}^a\\
=&\frac{1}{\sqrt{2}}\left(\varsigma_o (0,1,0,-i)+\bar\varsigma_o (0,1,0,i)\right)\\
=&\frac{1}{\sqrt{2}}\left(0,(\varsigma_o+\bar{\varsigma}_o),0,
i(\bar{\varsigma}_o-\varsigma_o)  \right)\\
=&\frac{2}{\sqrt{2}}\left(0,\varsigma_{oR},0,\varsigma_{oI}\right);
\end{split}
\end{equation} 
and a similar expression is obtained for $\varsigma^a_s$.

Therefore, by replacing into eq.(\ref{eq:deltabeta}), we obtain
\begin{eqnarray}
\varsigma_{sR}&=&(1-\kappa-\gamma_1 )\varsigma_{oR}-\gamma_2\,\varsigma_{oI},\\
\varsigma_{sI}&=&-\gamma_2\,\varsigma_{oR}+(1-\kappa+\gamma_1)\varsigma_{oI};
\end{eqnarray}
which has a direct physical interpretation in the plane of the image,
since the real component represents horizontal direction and
the imaginary component the vertical one.

The equation satisfied by $M$ is found from equation (\ref{eq: dev geo pert0}):
\begin{equation}
\begin{split} 
\ell\big( \ell\left(M D_A \boldsymbol{\mathcal{V}}_0 \right) \big) &= 
\ell\big( \ell\left(M\right) D_A \boldsymbol{\mathcal{V}}_0 + 
M \ell\left(D_A \boldsymbol{\mathcal{V}}_0 \right) \big) 
\\
& = \ell\big( \ell\left( M \right) \big)D_A \boldsymbol{\mathcal{V}}_0 + 
2\ell\left(M\right)\ell\left(D_A \boldsymbol{\mathcal{V}}_0 \right) \\
&+ M \ell\big( \ell\left( D_A \boldsymbol{\mathcal{V}}_0 \right) \big) \\
& = \left( \ell\big( \ell\left( M \right) \big) - 
2\ell\left(M\right)P_B  \right) 	D_A \boldsymbol{\mathcal{V}}_0 \\
&- MQ_B D_A \boldsymbol{\mathcal{V}}_0 \\
& = -\left(Q_B +  Q_L \right)M D_A \boldsymbol{\mathcal{V}}_0;
\end{split}
\end{equation}
which simplifies to
\begin{equation}
\ell\big( \ell\left( M \right) \big) - 2 \ell\left( M \right)P_B + Q_L\, M = 0.
\end{equation}
Now, using the fact that for R-W geometries one has the following 
important relation
\begin{equation}
P_B =  \mathbb{I}\, \rho = -\mathbb{I} \frac{\ell\left( D_A^2 \right)}{2 D_A^2};
\end{equation}
one arrives to the final equation:
\begin{equation}\label{eq: M}
\ell \big( D_A^2 \ell\left( M \right) \big) + Q_L \, D_A^2 M = 0.
\end{equation}

\subsection{Solving the matrix equation}\label{subsec:Solving-Matrix-M}
In order to proceed to the integration of equation (\ref{eq: M}) the initial conditions
must be specified. One can see that the appropriate ones for our problem are:
\begin{align}
\left( \left. D_A^2 \, \ell(M) \right) \right|_{{r}=0} &= 0, \\
\left. M \right|_{{r}=0} &= \mathbb{I} .
\end{align}

Using the first initial condition we obtain
\begin{equation}
{\ell}\left( M \right) = -\frac{1}{D_A^2}\int_{0}^{r} Q_L \, D_A^2 M \, dr';
\end{equation}
while in the next step we found 
\begin{equation}
M = \mathbb{I} - \int_{0}^{{r}_s} \left( \frac{1}{D_A^2} \int_{0}^{r'} Q_L
\, D_A^2 M \, dr'' \right)dr';
\end{equation}
where $r_s$ denotes the position of the emitting source.

Let us remember that the affine parameter $r$ takes negative values when it approaches to 
the source; due to this fact we prefer to use the normalized affine distance $\lambda$ to 
parametrize the geodesic. 
With this choice we rewrite the last equation to
\begin{equation}\label{eq:integral_M}
M = \mathbb{I} - \int_{0}^{\lambda_s} \left( \frac{1}{D_A^2} \int_{0}^{\lambda'}
Q_L \, D_A^2 M \, d\lambda'' \right) d\lambda'.
\end{equation}

Since we are only considering first order effects of $Q_L$ 
one can use iterations to solve this integro-differential
equations in orders of the curvature deviation. 
One can see that the leading order is
\begin{equation}\label{eq:M-first-order-double-intergal}
M = \mathbb{I} - \int_{0}^{\lambda_s} \left( \frac{1}{D_A^2}
\int_{0}^{\lambda'} Q_L \, D_A^2 \, d\lambda'' \right) d\lambda'.
\end{equation}

This is the main equation which describes the distortions at first order of isolated
distributions over the cosmological background here considered.

\subsection{Optical scalars}
It was mentioned before that the matrix $\mathcal{A}$ containing the optical scalars 
is given by:
\begin{equation}\tag{\ref{eq:Matrix-A}}
\mathcal{A} = \left(1 - \kappa_c \right)M.
\end{equation}
It will be used together with equation (\ref{eq:M-first-order-double-intergal}) 
to obtain the optical scalars. This task is most easily done working with the real and imaginary
parts of the deviation vector $\varsigma$; namely $\varsigma =\varsigma_{R} + 
i\varsigma_{I}$ and the real and imaginary parts of the Weyl curvature scalar, 
$\Psi^L_{0} =\Psi^L_{0R} + i \, \Psi^L_{0I}$.

One finds the following structure in the optical matrix
\begin{equation}\label{eq:acosmic}
\mathcal{A} = \left(1 - \kappa_c \right)
\begin{pmatrix}
1 - \kappa_L - \gamma_{1L} & -\gamma_{2L} \\
-\gamma_{2L} & 1 - \kappa_L + \gamma_{1L} 
\end{pmatrix};
\end{equation}
where the \emph{intrinsic optical terms} to the additional lens
$\left(\kappa_L, \gamma_{1L}, \gamma_{2L}\right)$ are explicitly
\begin{align}
\kappa_L &= \int_{0}^{\lambda_s} \left(\frac{1}{ D_A^2 }\int_{0}^{\lambda'}
\Phi_{00}^L \, D_A^2 \, d\lambda'' \right) d\lambda' , 
\label{eq:kappal}
\\
\gamma_{1L} &= \int_{0}^{\lambda_s} \left(\frac{1}{ D_A^2 }\int_{0}^{\lambda'}
\Psi_{0R}^L \, D_A^2 \, d\lambda'' \right) d\lambda' , 
\label{eq:gammal}\\
\gamma_{2L} &= \int_{0}^{\lambda_s} \left( \frac{1}{ D_A^2 }\int_{0}^{\lambda'}
\Psi_{0I}^L \, D_A^2 \, d\lambda'' \right) d\lambda'
\label{eq:gamma2} ;
\end{align}
where the matrix in (\ref{eq:acosmic}) refers to the physical
frame in the image plane.

The subindex $L$ is introduced at this moment to distinguish quantities depending on
the isolated inhomogeneities in contrast to background quantities such as $\kappa_c$.

Then, one can write the optical scalars for the whole system of an additional lens
regarded over a R-W spacetime:
\begin{align}
\kappa &= \left(1 - \kappa_c \right)\kappa_L + \kappa_c, \label{eq:kappa-general}\\
\gamma_1 &= \left(1 - \kappa_c \right)\gamma_{1L}, \label{eq:gamma-1-general} \\
\gamma_2 &= \left(1 - \kappa_c \right)\gamma_{2L}. \label{eq:gamma-2-general}
\end{align}		

One can see that the contribution of the background is two fold; 
on one hand, it is present in the intrinsic optical scalar trough the 
geometric distance and the area distance and, on the other hand it also appears 
by means of the modulating factor $\left( 1 - \kappa_c \right)$.

The above expressions are the generalization for the optical scalars of a lens in 
the standard cosmological context. 

In the simpler case of a Minkowski or a Milne background spacetime, where 
$D_A = \lambda$, one arrives at the expression already presented in \cite{Gallo11};
they are:
\begin{align}
\kappa &= \frac{1}{\lambda_s}\int_{0}^{\lambda_s} \lambda 
\left( \lambda_s - \lambda \right) \Phi_{00}^L \, d\lambda', \\
\gamma_1 &= \frac{1}{\lambda_s}\int_{0}^{\lambda_s} \lambda 
\left( \lambda_s - \lambda \right) \Psi_{0R}^L \, d\lambda', \\
\gamma_2 &= \frac{1}{\lambda_s}\int_{0}^{\lambda_s} \lambda
\left( \lambda_s - \lambda \right) \Psi_{0I}^L \, d\lambda';
\end{align}
in which the double integral has been expressed into a single one by means
of a integration by parts.

\section{The axially symmetric case }\label{sec:Axial-symmetry}

The expressions presented in the last sections do not involve
any assumption on the nature and geometry of the source; 
in this section, instead, we discuss models with axial symmetry, in which the axis 
of symmetry lies along the line of sight passing through the central region of 
the distribution. 
We are interested then in the study of a photon traveling along a direction
parallel to the axis of symmetry which has an
impact parameter $J$ with respect to the centre of the distribution. 
For this purpose we will take an adapted coordinate system to the situation in which
two mutually orthogonal directions $x$ and $z$ are considered.
They are also orthogonal to the direction $y$ in which the photons travel.
Due to the symmetry of the lens it will be also useful to work in terms of 
the angle $\vartheta$ between the $z-$axis and the path of the photon.

Then, with this setting one notes that the component $\Psi_{0}^L$ is a 
spin zero real quantity, and it depends on the $(J, \lambda)$ coordinates, while
the component $\Psi_{0}^L$ is a spin two complex quantity and it has 
the functional dependence
\begin{equation}
\Psi_{0}^L = |\Psi_{0}^L|e^{2i\vartheta + \text{phase}};
\end{equation}
where the phase is gauge dependent.

As it was mentioned in \cite{Gallo11}, it is usual to define the real quantity 
$\psi_{0}^L \left( J, \lambda \right)$ associated to the Weyl contribution from
\begin{equation}
\Psi_{0}^L \left( J, \lambda, \vartheta \right)= - 
\psi_{0}^L \left( J, \lambda \right)e^{2i\vartheta}.
\end{equation}

With this definition one writes for the intrinsic optical scalars of the lens:
\begin{align}
\kappa_L (J) &= \int_{0}^{\lambda_s} \left(\frac{1}{D_A^2} \int_{0}^{\lambda'}
\Phi_{00}^L(J, \lambda) D_A^2  d\lambda'' \right) d\lambda' , 
\tag{\ref{eq:kappal}}
\\
\gamma_{cL} (J, \vartheta) &= -e^{2i\vartheta}
\int_{0}^{\lambda_s} \left(\frac{1}{D_A^2}\int_{0}^{\lambda'}
\psi_{0}^L \left(J, \lambda \right) \, D_A^2 d\lambda'' \right)d\lambda';
\end{align}
where the complex shear is normally expressed in terms of its
real and imaginary parts, namely $\gamma_{cL} \equiv \gamma_{1L} + i \gamma_{2L}$.

The complex form for the shear components invite us to define the real quantity 
$\gamma_L \left(J \right)$ as 
\begin{equation}
\big( \gamma_{1L} + i \gamma_{2L} \big)(J, \vartheta) \equiv 
- \gamma_L\left(J \right)e^{2i\vartheta};
\end{equation}
so that one simply has
\begin{equation}
\gamma_L\left(J \right) = 
\int_{0}^{\lambda_s}{\left(\frac{1}{{D_A^2}}\int_{0}^{\lambda'}
	{\psi_{0}^L\left( J \right) {D_A^2} d\lambda''}\right)d\lambda'}.
\end{equation}

\section{The thin lens approximation}\label{sec:thin-lenses}
The common situation in the cosmological scenario is the one in which the
typical size $\delta_l$ of the lens is much smaller than
the distance from the observer to the lens $\lambda_l$, and
than the distance from the lens to the source $\lambda_{ls}$.
This configuration is referred to as a thin lens. 

Following the same lines  presented in article \cite{Gallo11} we will assume that
the   scalars of curvature $\left\lbrace \Phi_{00}^L, 
{\Psi_{0}^L} \right\rbrace$, denoted generically as ${\bar{C}^L}$ 
will be sharply peaked around $\lambda_l$, where the lens is located. 
This implies that the following approximation must to hold
\begin{equation}\label{eq:definiciondehat}
{\bar{C}^L}(\lambda) 
\equiv 
\int_{0}^{\lambda}{{\delta C}(\lambda') d\lambda'} 
\cong 
\begin{cases}
0, & \forall \, \lambda < \lambda_l -  \delta_l \\
{\widehat{C}^L}, & \forall \, \lambda \geq \lambda_l +  \delta_l
\end{cases};
\end{equation}
where $\delta_l \ll \lambda_l$, $ \delta_l \ll \lambda_s$ and $\delta_l \ll \lambda_{ls}$
($\lambda_{ls} = \lambda_s - \lambda_l$).

We make use of this assumption to simplify the integrand inside of parenthesis in the
expressions in equations (\ref{eq:kappal}, \ref{eq:gammal}, \ref{eq:gamma2}). 
Let us note that
\begin{equation}
\begin{split}
\frac{1}{{D_A^2}}\int_{0}^{\lambda} { {C^L} \, {D_A^2} d\lambda'} &= 
{\widehat{C}^L} - \frac{1}{{D_A^2}}\int_{0}^{\lambda}{\frac{d {D_A^2}}{d\lambda'}
	{{\bar{C}^L}}(\lambda') d\lambda'} \\
&= {\widehat{C}^L} - \frac{{\widehat{C}^L}}{{D_A^2}}
\bigg({{D_A^2}}  - D_A^2(\lambda_l) \bigg)  \\
&= {\widehat{C}^L} \frac{ D_A^2(\lambda_l) }{{D_A^2}};
\end{split}
\end{equation}
where it is clear that one must have $\lambda_l \leqslant \lambda$.
Then, one can take another integral to arrive at reduced expressions for the 
optical scalars; one finds 
\begin{equation}
\begin{split}
\int_{0}^{\lambda_s}{\left(\frac{1}{{D_A^2}(\lambda')}\int_{0}^{\lambda'}
	{{C^L} \,{D_A^2} d\lambda''}\right)d\lambda'} =
{\widehat{C}^L} D_A^2(\lambda_l)\int_{\lambda_l}^{\lambda_s}{\frac{d\lambda'}
	{{D_A^2}}};
\end{split}
\end{equation}
where we have neglected terms of order $\mathscr{O}\left( \frac{\delta_l}{\lambda_l}\right)$.

Since the factor in the last equation will appear recurrently we define, for 
the sake of simplicity in the notation, the symbol:
\begin{equation}\label{eq:Dls-factor}
\mathbf{D}_{ls} \equiv 
D_A^2(\lambda_l) \int_{\lambda_l}^{\lambda_s}{\frac{d\lambda'}
	{{D_A^2}(\lambda')}}.
\end{equation}

Then, the intrinsic optical scalars of the lens acquire a simpler form:
\begin{align}\label{eq:thin-optical-scalar}
%\Aboxed{ 
\kappa_L &= \mathbf{D}_{ls} {\widehat{\Phi}^L_{00}} 
%}
, \\
%\Aboxed{ 
\gamma_{1L} &= \mathbf{D}_{ls}{\widehat{\Psi}^L_{0R}} 
\label{eq:thin-optical-scalar2}
%} 
, \\
%\Aboxed{ 
\gamma_{2L} &= \mathbf{D}_{ls}{\widehat{\Psi}^L_{0I}}  
\label{eq:thin-optical-scalar3}
%}
; 
\end{align}
while the optical scalars of the whole system 
\begin{align}
%\Aboxed{
\kappa &= \left( 1 - \kappa_c  \right)
\mathbf{D}_{ls} {\widehat{\Phi}^L_{00}}  + \kappa_c \label{eq:kapa_total}
%}
, \\
%\Aboxed{
\gamma_{1} &= \left( 1 - \kappa_c  \right) \mathbf{D}_{ls} {\widehat{\Psi}^L_{0R}}  
\label{eq:gama1_total}
%}
, \\
%\Aboxed{
\gamma_{2} &= \left( 1 - \kappa_c  \right) \mathbf{D}_{ls} {\widehat{\Psi}^L_{0I}}  
\label{eq:gama2_total}
%}
;
\end{align}
where, from the above notation one is using:
\begin{equation}\label{eq:hatfi}
{\widehat{\Phi}^L_{00}} 
= \int_{0}^{\lambda_s} {\Phi^L_{00}}  d\lambda'
,
\end{equation}
and
\begin{equation}\label{eq:hatpsi}
{\widehat{\Psi}^L_{0}} 
= \int_{0}^{\lambda_s} {\Psi^L_{0}}  d\lambda'
.
\end{equation}
%\verde{
Notice that since $\kappa_c$ is order $\mathscr{O}(Q_B)$,
and $\kappa_L$ and $\gamma_L$ are order $\mathscr{O}(Q_L)$;
one finds terms of order order $\mathscr{O}(Q_B \, Q_L)$
in equations (\ref{eq:kapa_total})-(\ref{eq:gama2_total}).
%	}

In section \ref{subsec:Examples} we will show the connection of the above expressions
with usual formulae present in the literature.
For the moment, let us note that,
in the limit of no cosmological background one has that $D_A$ converges to $\lambda$ and
\begin{equation}\label{eq:Dls-to-standard-quotient}
\mathbf{D}_{ls} \rightarrow \frac{\lambda_l \lambda_{ls}}{\lambda_s}
;
\end{equation}
so that in this way one arrives at the usual expressions found in the studies of
gravitational lenses over a flat background\cite{Gallo11}.

It is worthwhile to remark that the expressions 
(\ref{eq:thin-optical-scalar})-(\ref{eq:thin-optical-scalar3})
are built out of two distinctive factors; namely the $\textbf{D}_{ls}$ and the
hated quantities that are calculated from the curvature of the lens.
The first factor does not depend on the possible motion of the lens, and it only
depends on the cosmological scenario. Instead the hated quantities depend
on the possible motion of the lens; which is unavoidable in the cosmological context.
In the next two subsection we concentrate on both factors.

\subsection{Alternative expression for the factor $\textbf{D}_{ls}$}

The term, $\textbf{D}_{ls}$, which only contains information about the cosmology, can be 
written in a more clarifying form using the coordinates $(u,\chi)$ discussed in section
\ref{subsec:inertial-u}. 
In fact, we show that the integrand appearing in equation (\ref{eq:Dls-factor}) 
is a total derivative along the null geodesic.

Let us note that from equation (\ref{eq:dr-dchi-du}) and the fact that 
$\frac{\partial}{\partial \lambda} = -\ell$ one has the following relation 
\begin{equation}
d\lambda = \frac{A^2(u,\chi)}{A(t_0)}  d\chi;
\end{equation}
which means that
\begin{equation}\label{eq:Dls-simplificado}
\begin{split}
\textbf{D}_{ls} &=
\frac{D_A^2(\chi_l)}{A(t_0)}\int_{\chi_l}^{\chi_s}\frac{d\chi}{f_k^2(\chi)} \\
&= \frac{D_A^2(\chi_l)}{A(t_0)} \frac{f_k(\chi_s- \chi_l)}{f_k(\chi_s)f_k(\chi_l)} \\
&= \frac{1}{\left( 1 + z_l \right)}\frac{D_{A}(\chi_s - \chi_l)  D_{A}(\chi_l) }{D_{A}(\chi_s)}
;
\end{split}
\end{equation}
where it must be understood that $z_l$ is the redshift at the place of the lens and 
$D_{A}(\chi_s - \chi_l)$ is the angular diameter distance measured by a fundamental observer 
at the coordinates $\left(u, \chi_l \right)$ 
\begin{equation}\label{eq:D_Als}
D_{A}(\chi_s - \chi_l ) =  A(u, \chi_s ) f_k(\chi_s - \chi_l).
\end{equation}
In what follows we will use a short notation in which we only retain
the subindex of the coordinate $\chi$, namely:
\begin{align}\label{eq:DLS}
\textbf{D}_{ls} &= \frac{1}{1 + z_l}\frac{D_{A_{ls}} D_{A_{l}}}{D_{A_s}}.
\end{align}

Let us observe that the presence of the redshift factor $1 + z_l$ could be source of 
confusion if one consider the case of a Milne (vacuum) spacetime where 
$D_A = \lambda$ but $z_l \neq 0$ since in this case it appear that we will not
recover the expression (\ref{eq:Dls-to-standard-quotient}).
This is just apparent since in this case one has that equation (\ref{eq:D_Als})
becomes $(1 + z_l)(\lambda_s - \lambda_l)$ as one can easily check.
This is due to the fact that the area distance depends on the motion of the
observer which in this case correspond to the comoving cosmic observer. 
In other words, one has for Milne spacetime
\begin{equation}
D_{A_{ls}} \to \bar{\lambda}_s - \bar{\lambda}_s = (1 + z_l)(\lambda_s - \lambda_l);
\end{equation}
where $\bar{\lambda}$ refers to the affine parameter defined by equation (\ref{eq:vconele}) 
respect to the worldline of the comoving cosmic observer.

\subsection{Integration on the curvature for moving lenses}

In principle, expressions (\ref{eq:hatfi}) and (\ref{eq:hatpsi}) have all the information
one needs to complete the calculation of the optical scalars.
But in the case the lens is moving one might like to refer the calculation to
its intrinsic rest frame. To do this one can think in leaving the null frame of the
observer fixed an calculate the curvature of moving sources, or 
one can think in leaving the geometry unaffected
and only change the frame of observation by the appropriate boost.
From the GHP formalism\cite{Geroch73} one knows
that both integrands have boost weight 2.
Then, taking into account the change in the affine parameter for the boosted frame,
one deduces that for a local stationary piece of of spacetime which is moving relative
to the observer with four velocity $v^a$, the expressions (\ref{eq:hatfi}) and (\ref{eq:hatpsi})
are related with respect to the non-moving case by a factor
\begin{equation}
\frac{1}{\ell \cdot v}
= 1 + z_v .
\end{equation}
Let us emphasize that $z_v$ does not need to agree with $z_l$.

Then, one can see that expressions (\ref{eq:thin-optical-scalar})-(\ref{eq:thin-optical-scalar3}),
have all the same structure; namely,
if $\mathsf{o}_{Lv}$ is the optical scalar for a lens with velocity $v$,
one can express
\begin{equation}\label{eq:olv-ol}
\begin{split}
\mathsf{o}_{Lv} &=  \textbf{D}_{ls}  {\widehat{C}^L_v}   \\
&=  \textbf{D}_{ls} (1 + z_v)  {\widehat{C}^L}   \\
&=   \frac{(1 + z_v)}{(1 + z_l)} \frac{D_{A_{ls}} D_{A_{l}}}{D_{A_s}} {\widehat{C}^L} \\
& =  \frac{(1 + z_v)}{(1 + z_l)} \mathsf{o}_{L} ;
\end{split}
\end{equation}
where we are using the notation ${\widehat{C}^L_v}$ for the moving lens
and ${\widehat{C}^L}$ for the lens at rest with respect to the observer.

The simplicity of equation (\ref{eq:olv-ol}), and in its derivation, should not hide that
this is a noteworthy expression that synthesizes all that one needs to
take into account in the motion of the lenses; therefore in this way
we simplify previous works\cite{Kopeikin:1999ev,Frittelli:2002yx,Wucknitz:2004tw} 
on the subject of
moving gravitational lenses, and generalized its application to the cosmological
scenario.

%\paragraph{A note regarding $\Sigma_{cr}$:}
{\bf A note regarding $\Sigma_{cr}$:}

At this point we would also like to take the opportunity to make a remark
regarding the notion of critical mass density used in several previous works and textbooks.

We have noticed before that the contribution to the optical gravitational lens scalars
appearing in (\ref{eq:thin-optical-scalar})-(\ref{eq:thin-optical-scalar3}) are composed
of two distinctive factors involving direct contributions from the cosmological
background and the lens respectively.
When comparing with the textbook\cite{Schneider92, Schneider06} recipe based 
on the $\Sigma/\Sigma_{cr}$ approach,
we should relate our $\textbf{D}_{ls}$ with $1/\Sigma_{cr}$
and ${\widehat{C}^L_v}$ with  $\Sigma$.
Let us point out that
in a strict sense,
we have found that $\textbf{D}_{ls}$ does not 
coincides with $1/\Sigma_{cr}$ 
as defined by\cite{Schneider06}
\begin{equation}
\frac{1}{\Sigma_{cr}} \equiv 
\frac{4\pi G}{c^2}
\frac{D_A(\chi_s - \chi_l) D_A(\chi_l)}
{D_A(\chi_s)} ;
\end{equation}
since, up to the numerical factor with the physical constants, 
%which in our results is contained
%in the integrals (\ref{eq:hatfi}), (\ref{eq:hatpsi}); 
we note that the factor 
involving the redshift of the lens is missing.
This is important because in reference \cite{Schneider06} it has been claimed that to
do calculations in the cosmological context, one `only' needs to replace the
flat area distances with the cosmological ones.

%\paragraph{A note regarding $\Sigma$:}
{\bf A note regarding $\Sigma$:}

Furthermore, in the textbook \cite{Schneider92}, near their equations (4.19) and (4.20), it
is calculated that the deflecting angle is unaffected, by first order effects
on the velocity of the lens. Then since the quantity $\Sigma$ used in these references
is just linear in the deflecting angle, one would deduce that  $\Sigma$
is unchanged by the motion of the lens.
We have shown above that this is not correct.

This means that it is wrong to generalize the common expressions appearing in 
weak lensing studies in the case of a flat background to the cosmological scenario,
just by changing distances to angular diameter distances.	
We will return to this point below when presenting basic examples.

\section{Stationary spherically symmetric moving thin lenses}\label{sec:esferica-lenses}

Let us consider the case in which locally, the gravitational lens, is represented
by a stationary spherically symmetric line element.
Then, in a neighbourhood of the lens, we can set a coordinate system where
the line element of the geometry can be expressed as:
\begin{equation}\label{eq:ds-esferico}
ds^2_L = e^{2\Phi(\mathbf{r})} d\mathbf{t}^2 - \frac{d\mathbf{r}^2 }
{1 - \frac{2  M(\mathbf{r})}{\mathbf{r}}} - \mathbf{r}^2\left( d\theta^2 + 
\sin^2(\theta) d\phi^2 \right);
\end{equation}
which is completely determined by the functions $M(\mathbf{r})$ and $\Phi(\mathbf{r})$.

It is important to emphasize that the thin lens approximation as given in
expressions (\ref{eq:thin-optical-scalar})-(\ref{eq:thin-optical-scalar3}) 
can be applied to this situation. 
When this is done one notes also that the information 
provided by the underlined cosmology will appear  only 
trough the factor (\ref{eq:DLS}) containing the angular diameter 
distances and the redshift.

Furthermore, comparing the above equations (\ref{eq:thin-optical-scalar})-(\ref{eq:thin-optical-scalar3})
with equations (78)-(80) of reference \cite{Gallo11}, and realizing that in the 
thin lens approximation, what matters is the impact parameter $J$ from the center;
one can note that all the discussion of section V of \cite{Gallo11} can be applied
also to the cosmological scenario; taking into account the motion of the lens
discuss previously.
This means that we can express the optical scalars in terms 
of the energy-momentum components and the function $M(\mathbf{r})$ as:
\begin{equation}\label{eq:kapaesferica}
\begin{split}
\kappa_L\left( J \right) =& 
\frac{4\pi G}{c^2} \mathbf{D}_{ls} (1 + z_v)
\int_{-\infty}^{\infty} \bigg[ \varrho(\mathbf{r}) + \frac{ P_r(\mathbf{r})}{c^2} \\
&+ 
\frac{J^2}{c^2 \mathbf{r}^2}\bigg(  P_t(\mathbf{r}) -  P_r(\mathbf{r}) \bigg) \bigg] dy 
,
\end{split}
\end{equation}
and
\begin{equation}
\begin{split}
\gamma_L\left(J \right) =& 
\frac{G}{c^2}\mathbf{D}_{ls} (1 + z_v)
\int_{-\infty}^{\infty} \frac{J^2}{\mathbf{r}^2}\bigg[ \frac{3 M(\mathbf{r})}{\mathbf{r}^3} \\
&- 
\bigg( \varrho(\mathbf{r}) + \frac{ P_t(\mathbf{r})}{c^2} 
- \frac{ P_r(\mathbf{r})}{c^2} \bigg) 
\bigg] dy ; \label{eq:gamaesferica}
\end{split}
\end{equation}
where  $\varrho$ denotes the energy density of the distribution 
and $ P_r(r)$ and $ P_t(r)$
are the spacelike components of the energy-momentum tensor.
In order to simplify the notation we are omitting
{the upper indices `L'} in the curvature expressions.
The integration variable $y$ is defined as in \cite{Gallo11}:
\begin{equation}
\mathbf{r}^2 \equiv J^2 + y^2
.
\end{equation}

It is probably worthwhile noting that the standard assumption in cosmological
studies is to neglect the spacelike components of the energy-momentum tensor.

Equations (\ref{eq:kapaesferica}) and (\ref{eq:gamaesferica}) are very useful expressions
which allow us to write in a simple manner a relation between
the optical scalars of the gravitational lens and the matter content;
and constitute the generalization of the pair of equations (183) of \cite{Gallo11} 
to the cosmological context.

In the static spherically symmetric lens discussed above, we have assumed that
locally the geometry has this behavior so that we can use the 
standard line element, where the curvature need not be small
with respect to the cosmological background.
However one could also deal with the situation in which one has a
local static spherically symmetric distribution of matter which is small
with respect to the local cosmic energy density. In this situation,
we can still use the above expressions, now with the understanding
that the quantities refer with respect to the cosmological background.
So that, for example, in the discussion of the energy density,
if we denote with $\varrho_c$ the cosmological energy density corresponding 
to the position of the lens, one would use  ${\varrho^L}$
instead of $\rho$ in the above expressions, and would have
${\varrho^L} = \varrho_{\text{proper}} - \varrho_c$.

\subsection{Examples}\label{subsec:Examples}
In this subsection we apply the generalized expressions for the optical
scalars of a monopole mass, for the well known isothermal mass 
density distribution, and 
we also present a peculiar geometry of lens. 
We take the opportunity to point out an 
interesting relation that this unusual distribution shares with the 
isothermal one.

\subsubsection{A monopole mass (Schwarzschild) }

Let us consider here a monopole mass characterized by a mass $M=$constant; which
it could be moving with velocity $v$ with respect to the observer. 
For a such lens one finds that
\begin{align}
\kappa_L &= 0, \\
\gamma_L &= \frac{4G}{c^2} \mathbf{D}_{ls}  (1 + z_v) \frac{M}{ J^2};
\end{align}
then the total convergence and shear are:
\begin{equation}
\kappa = \kappa_c 
,
\end{equation}
\begin{equation}
\begin{split}
\gamma &= (1 - \kappa_c )  \frac{4G}{c^2} \mathbf{D}_{ls} (1 + z_v)\frac{M}{J^2} \\
& = \frac{4G}{c^2}
\frac{(1 - \kappa_c )}{\left( 1 + z_l \right)}\frac{D_{A_{sl}}  D_{A_l} }{D_{A_s}}
(1 + z_v) \frac{M}{J^2}.
\end{split}
\end{equation}

\subsubsection{The isothermal profile}
The isothermal profile is characterized by a mass density
function of the form
\begin{equation}\label{eq:rho-isothermico}
\varrho(\mathbf{r}) = \frac{\sigma^2}{2 G \pi \mathbf{r}^2}.
\end{equation}

This distribution
appears in many astrophysical studies in which it is often used as 
profile for stellar dynamics models and for the study of the differential rotation of 
galaxies whit dark matter among other applications.
The parameter $\mathsf{v} \equiv \sqrt{2} \sigma$ has unit of velocity and it is assumed to 
satisfy $\mathsf{v} \ll c$; which implies that pressures in this model can be neglected.
This characteristic velocity of the isothermal profile should not be confused with the
possible velocity of the lens $v$.
Then,  the matter content is characterized by
\begin{align}
M(\mathbf{r}) &= \frac{2 \sigma^2}{G} \mathbf{r}, \\
P_r(\mathbf{r}) &= P_t(\mathbf{r}) \approx 0;
\end{align}
together with equation (\ref{eq:rho-isothermico}). 

It yields the following intrinsic optical scalar for the lens:
\begin{align}
\kappa_L &= \frac{\mathbf{D}_{ls}}{c^2} (1 + z_v) \frac{2 \pi \sigma^2 }{J}, \\
\gamma_L &= \frac{\mathbf{D}_{ls}}{c^2} (1 + z_v) \frac{2 \pi \sigma^2 }{J};
\label{eq:shearL-isoth-profile}
\end{align}
and if we include
the proper contribution from the background we obtain:
\begin{align}
\begin{split}
\kappa &= (1 - \kappa_c ) \frac{\mathbf{D}_{ls}}{c^2} (1 + z_v) 
\frac{2 \pi \sigma^2 }{J} + \kappa_c \\
& = \frac{(1 - \kappa_c )}{\left( 1 + z_l \right)}\frac{D_{A_{sl}}  D_{A_l} }{c^2 D_{A_s}}
(1 + z_v) \frac{2 \pi \sigma^2 }{J} + \kappa_c , 
\end{split}
\\
\begin{split}
\gamma &=  (1 - \kappa_c ) \frac{\mathbf{D}_{ls}}{c^2} (1 + z_v) \frac{2 \pi \sigma^2 }{J} \\
& = \frac{(1 - \kappa_c )}{\left( 1 + z_l \right)}\frac{D_{A_{sl}}  D_{A_l} }{c^2 D_{A_s}}
(1 + z_v) \frac{2 \pi \sigma^2 }{J}.
\end{split}
\end{align}

\subsubsection{Peculiar anisotropic solution}
The last example is an exact solution of the Einstein equation
which describes very well the phenomenology of the dark matter in astrophysical 
systems which was presented in \cite{Gallo:2011hi}. 
It posses a non-conventional energy-momentum tensor whose components are:
\begin{align}
\varrho(\mathbf{r}) &= 0, \\
P_r(\mathbf{r}) &= \frac{c^4}{4\pi G \mathbf{r}^2 \ln{\left( \frac{\mathbf{r}}{\mu}\right)}}, \\
P_t(\mathbf{r}) &=0;
\end{align}
with $\mu$ a constant of the distribution.
The solution satisfies the weak and strong energy conditions and it was
shown that the total mass of the distribution $M(\mathbf{r})$ vanishes.

In different application of this geometry to problems involving dark matter
it was found that in the range of interest $\ln \mathbf{r}$ was much smaller
than $-\ln \mu$. This means that for any practical purposes, the $log$ factor could 
be identified with a constant;
in other words one could take $4\Delta^2 \cong \frac{c^2}{\ln{\left( \frac{\mathbf{r}}{\mu}\right)}}$.
Then, by using equations (\ref{eq:kapaesferica}) and (\ref{eq:gamaesferica}) one arrives at:
\begin{align}
\kappa_L \left( J \right) &= \frac{\mathbf{D}_{ls}}{c^2} (1 + z_v) \frac{2\pi \Delta^2}{J}, \\
\gamma_L \left( J \right) &= \frac{\mathbf{D}_{ls}}{c^2} (1 + z_v) \frac{2\pi \Delta^2}{J}.
\end{align}
Therefore we see that this type of models behave as an isothermal profile with velocity 
dispersion $\mathsf{v} =  \sqrt{2} \Delta$ when it is regarded as a gravitational lens.
The shear and the convergence are
\begin{align}
\begin{split}
\kappa &= (1 - \kappa_c ) \frac{\mathbf{D}_{ls}}{c^2} (1 + z_v) 
\frac{2 \pi \Delta^2 }{J} + \kappa_c \\
& = \frac{(1 - \kappa_c )}{\left( 1 + z_l \right)}\frac{D_{A_{sl}}  D_{A_l} }{c^2 D_{A_s}}
(1 + z_v) \frac{2 \pi \Delta^2 }{J} + \kappa_c , 
\end{split}
\\
\begin{split}
\gamma &=  (1 - \kappa_c ) \frac{\mathbf{D}_{ls}}{c^2} (1 + z_v) \frac{2 \pi \Delta^2 }{J} \\
& = \frac{(1 - \kappa_c )}{\left( 1 + z_l \right)}\frac{D_{A_{sl}}  D_{A_l} }{c^2 D_{A_s}}
(1 + z_v) \frac{2 \pi \Delta^2 }{J}.
\end{split}
\end{align}

It is probably valuable to remark that if one were to use the textbook recipe,
based on the $\Sigma/\Sigma_\text{cr}$ quantities, one would obtain zero for all the optical
scalars in this case, since $\Sigma=0$.

%\paragraph{A note on the examples}
{\bf A note on the examples:}

These examples give the opportunity to show the main differences of our general
expressions with the paradigm based on the $\Sigma/\Sigma_\text{cr}$ approach.
The first two examples show that our expressions include the cosmic contribution
by the factor involving the cosmic convergence gravitational lens scalar $\kappa_c$
and the contribution of the general velocity of the lens by the factor 
involving the redshift $z_v$.
For lenses which are comoving with the fundamental observers of the background
R-W frame, one has a cancellation of this factor with the denominator involving
the cosmic background redshift; but the contribution from the 
the cosmic convergence gravitational lens scalar remains, so that even in
this case, our results do no agree with the simplified model based
on the $\Sigma/\Sigma_\text{cr}$  recipe.

In standard presentations of the above mentioned paradigm
the issue of possible motion of the lens
is completely  neglected, so that they omit the two 
effects that produces the two factors shown above;
and therefore their cancellation, for the case of a comoving lens,
has gone unnoticed in previous works.

Notably the last example demonstrate the inability of the mentioned paradigm to deal
with geometries that have been shown to be useful for the description
of the dark matter phenomena.

\section{Final comments}\label{sec:final-comments}

In this work we have presented a detailed systematic study of gravitational
lens optical scalars in the cosmological context.
We have also included new expressions for them, 
which allow for a general 
energy-momentum content of the lens and at the same time they contain the corrections
due to the motion of the lens.  
We also have presented new 
formulae for the relation of the observed magnitudes with the optical scalars 
through the intensity magnifications we define here.

In our approach we follow the usual framework for the study of gravitational
lens effects, in which one compares the lensed situation with the unlensed one;
in which no curvature is present.
In order to take into account the subtleties appearing in the discussion of 
lenses over a curved spacetimes, such as R-W, it was necessary to review the basic
language and interpretation of observations; which has been done
in section \ref{sec:Grav-lens}.
In particular we have remarked that the natural and universal notion
of distance, that can be applied to any spacetime,
is the affine distance $\lambda$.

It is interesting to note that
the whole R-W geometry can be characterized as a gravitational lens in its own right; 
which is thick, only producing convergence but not shear. 
It is also worthwhile to remark that our expressions for the cosmic convergence are valid for 
arbitrary large observing angles.

We have studied also the existence of an additional lens in a R-W scenario,
and have presented the new general expressions for the optical scalars, namely
equations (\ref{eq:kappal})-(\ref{eq:gamma-2-general}).
One should note that in several works sometimes
the effects of additional lenses are presented only with respect
to a R-W background, which corresponds only to equations 
(\ref{eq:kappal})-(\ref{eq:gamma2}); but in 
this case one would miss some effects of cosmological origin, 
that we include here explicitly.

Although there are excellent textbooks\cite{Schneider92, Schneider06} in the literature
that present a complete view of the subject of gravitational lens in a clear
pedagogical way, they rely on simplifying assumptions that severely
limit their use in detailed research of general astrophysical situations.
In particular they assume a typical lens is of Newtonian nature and not moving.
So, our
new expressions possess the interesting feature that they are not restricted to 
Newtonian like distributions of massive scatterer, nor are confined to be at rest; 
instead, they allow for a very general class of energy-momentum 
tensors which make 
these formulas a useful tool to address some of the difficulties
related to the missing mass problem in the Universe.
In this way we generalize expressions appearing in textbooks\cite{Schneider92, Schneider06},
reviews\cite{Wambsganss98, Bartelmann:1999yn, Bartelmann10}
and premises of research articles\cite{Kling:2007pw, Liao:2015uzb, Holanda:2015zpz}.

%\verde{
It is worthwhile to remark that our expressions for the gravitational lens
optical scalar are presented in terms of curvature components;
that is, prior to the choice of the field equation one would like
to use. But in all the examples we have discussed, we have used
the Hilbert-Einstein field equations.
%}

The thin lens approximation has proven\cite{Frittelli:2011uh} to be 
a good working hypothesis for a variety of systems.
Surprisingly, when the thin lens approximation is considered we have 
noticed the appearance of a factor involving the redshift of the lens;
as is indicated in equation (\ref{eq:Dls-simplificado});
which has not been considered previously. 
This means that the straightforward change of angular diameter distance,
used in the case of flat background,
to the angular diameter distance used on a R-W background,
is not enough to obtain the complete expression.
We remark this because several authors have suggested this wrong
technique, as for example in reference \cite{Schneider06};
where the factor involving the redshift is missing.
In order to remark the conceptual difference of our approach to those
found in standard textbooks, we have presented the general situation
in which the lenses might be moving with arbitrary velocity. This in turn
includes another redshift factor associated to the motion of the lens
which of course need not coincide with the velocity of the assumed
fundamental cosmic observers of a R-W background.

These equations are additionally useful and relevant to works 
concerning test of fundamental geometric relations in observational cosmology;
which are based on the use of gravitational 
lensing\cite{Holanda:2015zpz, Liao:2015uzb}.

We have considered few classic examples of
static and spherically symmetric lenses,
and applied the equations for the gravitational
lens optical scalars.
This is easily done, since we have provided with
expressions that relate the optical scalars
to the matter content of the lens.

It is probably worthwhile to point out that in our presentation we have not used
the notion of bending angle at all; since it is not essential, but it is useful
to show the difference with the standard approach based on this concept.
For those that would like to read a discussion of both concepts
for the case of gravitational lens with symmetries, we refer to our 
previous article \cite{Gallo11}. 
In relation to this, let us also mention that in the $\Sigma/\Sigma_\text{cr}$
paradigm, which is based in the bending angle concept, one can not discuss
the cosmic convergence $\kappa_c$ appearing in R-W spacetimes; 
since there is no bending of the light rays in this case (See also \cite{Pyne1996ApJ}). 
For this reason within this paradigm there appear no discussions of
the observations
of supernovae; instead we have shown here that the luminosity observations
of a supernova can be expressed in the terms of the gravitational lens 
optical scalar of the cosmological spacetime.

The connection with the observed magnitudes is done with
our introduction of
the intensity magnification $\tilde{\mu}$, defined in terms
of the fluxes, and we have also provided with an
expression of the \emph{distance modulus} $m - M$,
in terms of it, given in equations (\ref{eq:m-M}) and (\ref{eq:m-M_2}).
In order to make contact with the way in which the community describes the
behavior of the distance modulus, we have also introduced
an astrophysical cosmic
intensity magnification $\mu'_c$ that is thought as a function of redshift.
One can also express the distance modulus in terms of this magnification
as we have done in (\ref{eq:m-M_astro}).
It is important to remark that the physical intensity magnification $\tilde{\mu}$
turns out to depend only on the affine distance $\lambda$; and that
it coincides with the gravitational lens magnification $\mu$.
Therefore the distance modulus as expressed in equation (\ref{eq:m-M_2}),
valid in a general spacetime,
only depends on the gravitational lens scalars and kinematical data.
Instead, when expressing the distance modulus in terms of the
astrophysical  cosmic intensity magnification $\mu'_c$, 
it requires the assumption of a cosmological model, so that a
distance-redshift relations can be applied.
It is for this reason that something very peculiar happened with the way
in which the data of supernovae has been studied, namely:
one can see from equations (\ref{eq:m-M_2}) and (\ref{eq. dev geo}) that,
since in the calculation of the gravitational lens optical scalars
only intervene the traceless part of the Ricci tensor and the Weyl
tensor, the observations of the supernovae fluxes do not depend
on a possible cosmological constant, which only affects the
trace of the Ricci tensor. 
However in celebrated works, as
\cite{Riess:1998cb,Perlmutter:1998np}, 
the authors have studied the
relation between the behavior of the supernovae fluxes
with redshift, by the use of equation (\ref{eq:m-M_astro})
with the assumption of a cosmological model,
and argued that a cosmological constant
explains the observations.
The peculiarity is also related to this:
from tables \ref{tab:Planck-2} and \ref{tab:Baryonic} one can see that
the physical intensity magnification $\tilde \mu$ is always greater
than one; so that if one applies (\ref{eq:m-M_2})
to describe the observed luminosity supernovae, 
one would find that they are brighter than in the flat case; 
instead, in the
analysis of \cite{Riess:1998cb,Perlmutter:1998np} the claim is that
since the observed luminosity of supernovae is fainter 
(See table \ref{tab:Planck-2}) than expected as a function of redshift, 
then a cosmological constant would account for it.

It should be noticed that our arguments leading to the new expression
(\ref{eq:m-M_2}), should not be confused with the standard arguments
involving the statements of constant surface brightness as appear in
textbooks\cite{Schneider92}. Our argument relies on the validity of
the Etherington theorem, which is applicable to general spacetimes
and classes of lenses.

Let us note that in our direct approach to the study of gravitational
lens scalars, the discussion of moving lenses is straight forward and
it can be done in few lines, as expressed in the deduction of
equation (\ref{eq:olv-ol}). Instead in previous approaches based
on the bending angle concept, the discussions require much more
complicated algebra\cite{Kopeikin:1999ev,Frittelli:2002yx,Wucknitz:2004tw}.

Within the approximation of thin lenses we have considered configurations 
with a single 
lens but the treatment can easily be extended to complex 
arrangements of several lenses.
That is, one could consider having $N$ lenses, placed at cosmological 
distances $\lambda_1, \ldots, \lambda_N$, and apply the techniques
explained above.
We will tackle this problem in a further work.

\section*{Acknowledgements}

We have benefited from comments from an anonymous Reviever.
We acknowledge support from CONICET, SeCyT-UNC and Foncyt.

%%%%%%%%%%%%%%%%%%%%%%%%%%%%%%%%%%%%%%%%%%%%%%%%%%



\bibliographystyle{mnras}

\appendix

\section{Geometry in the coordinate systems adapted to the past 
	light cone}\label{ap:Geom-Null}

In this appendix 
we present a list of the geometric quantities that we have
employed for the description of the R-W spacetime in terms of the GHP 
formalism \cite{Geroch73}. 
We have taken the most relevant null tetrad for our purpose, 
namely those presented in subsection (\ref{subsec:inertial-u}).
For simplicity we present the expressions in terms of comoving 
coordinates.

Since Robertson-Walker geometries are conformally flat, its geometry is
characterized only by its connection and the Ricci curvature since
its Weyl tensor vanishes. 

\subsection{The connection and the curvature for the null tetrad associated to the null function $u$}

\subsubsection{Connection in terms of spin coefficients}

\begin{align}
\rho  &= - \frac{A(t_0)}{A(t)}\left( 
\frac{H(t)}{c} - \frac{\sqrt{1 - k f_k^2(\chi)}}{A(t) f_k(\chi)}\right)
,\\
\rho'  &= -\frac{A(t)}{A(t_0)} \left( \frac{H(t)}{c} + 
\frac{\sqrt{1 - k f_k^2(\chi)}}{ A(t) f_k(\chi)} \right)
,\\
\beta  &= -i \frac{\sqrt{2}}{4}\frac{\zeta}{A(t) f_k(\chi)}
,\\
\beta'  &= i \frac{\sqrt{2}}{4}\frac{\bar{\zeta}}{A(t) f_k(\chi)}
,\\
\epsilon'  &= \frac{A(t)}{2A(t_0)}\frac{H(t)}{c}
.
\end{align}

\subsubsection{Ricci curvature scalar}

\begin{align}
\Phi_{00} &= \frac{A^2(t_0)}{c^2 A^2(t)}
\left(\frac{k c^2}{A^2(t)} + H^2(t) + q(t)H^2(t) \right),\\
\Phi_{11} &= \frac{1}{4}\frac{A^2(t)}{A^2(t_0)}\Phi_{00}, \\
\Phi_{22} &= \frac{1}{4}\frac{A^4(t)}{A^4(t_0)}\Phi_{00} = 
\frac{A^2(t)}{A^2(t_0)}\Phi_{11},\\
\Lambda_{\mathtt{GHP}} &= \frac{1}{4c^2}\left( \frac{k c^2}{A^2(t)} + H^2(t) - q(t)H^2(t) \right).
\end{align}
Some of these equations are found in \cite{Moreschi:1988ik}

% Don't change these lines
\bsp	% typesetting comment
\label{lastpage}
\end{document}